\documentclass[preprint,aps]{revtex4}
\pdfoutput=1
\usepackage{dcolumn}% Align table columns on decimal point
\usepackage{bm}% bold math
\usepackage{graphicx}% Include figure files
\usepackage{epsfig}
\usepackage{amsmath}
\usepackage{amssymb}
\usepackage{mathrsfs}
\usepackage{ulem}
\usepackage{color}
\usepackage{hyperref}
\usepackage{float}
\allowdisplaybreaks

\def\fun#1#2{\lower3.6pt\vbox{\baselineskip0pt\lineskip.9pt
  \ialign{$\mathsurround=0pt#1\hfil##\hfil$\crcr#2\crcr\sim\crcr}}}

\def\simlt{\stackrel{<}{{}_\sim}}

\input epsf

\newenvironment{Eqnarray}%
         {\arraycolsep 0.14em\begin{eqnarray}}{\end{eqnarray}}
\newcommand{\be}{\begin{equation}}
\newcommand{\ee}{\end{equation}}
\newcommand{\bea}{\begin{Eqnarray}}
\newcommand{\eea}{\end{Eqnarray}}

\def\half{\tfrac{1}{2}}

\def\sinbii{s^2_\beta}
\def\sinbiv{s^4_\beta}
\def\cosb{c_\beta}
\def\sinb{s_\beta}

\def\cosbii{c^2_\beta}
\def\cosbiv{c^4_\beta}
\def\cbma{c_{\beta-\alpha}}
\def\sbma{s_{\beta-\alpha}}

\begin{document}

\begin{flushright}
\vspace{-0.2cm}EFI-18-1\\
\end{flushright}

%\title{Constraints on CP Violation in the MSSM Higgs Sector}
\title{ Wrong Sign Bottom Yukawa in Low Energy Supersymmetry }

\vspace*{0.2cm}

\author{
Nina M. Coyle$^a$, Bing Li$^a$ and Carlos E.M. Wagner$^{a,b,c}$
\vspace{0.2cm}
\mbox{}
 }
\affiliation{
\vspace*{.5cm}
$^a$~\mbox{Physics Department and Enrico Fermi Institute, University of Chicago, Chicago, IL 60637}\\
$^b$  \mbox{Kavli Institute for Cosmological Physics, University of Chicago, Chicago, IL 60637}\\
$^c$ \mbox{High Energy Physics Division, Argonne National Laboratory, Argonne, IL 60439}\\
}

\begin{abstract}
The study of the Higgs boson properties is one of the most relevant activities in current particle physics.  In particular,
the Higgs boson couplings to third generation fermions is an important test of the mechanism of mass generation.
In spite of their impact on the production and decay properties of the Higgs boson, the values of these  couplings
are still uncertain and, in models of new physics, they can differ in magnitude as well as in sign with respect to the
Standard Model case. In this article, we study the  possibility of a wrong sign bottom-quark Yukawa coupling within
the framework of the Minimal and Next-to-Minimal
Supersymmetric Standard Model. Possible experimental tests are also discussed, including novel decays of the
heavy CP-even and CP-odd Higgs fields that may be probed in the near future and that may lead to an explanation
of some intriguing di-boson signatures observed at the ATLAS experiment. 

\end{abstract}
\thispagestyle{empty}

\maketitle

\section{Introduction}

The Higgs discovery~\cite{Aad:2012tfa},\cite{Chatrchyan:2012ufa} has led to the confirmation of  the Standard Model
as the proper effective theory at the weak scale.
No new particle has been seen at the LHC, implying that new physics at the weak scale should be weakly interacting or that
strongly interacting particles, if present, should lead to signatures involving soft decay products or in channels with large
irreducible backgrounds.  Searches for new physics under these conditions should be complemented by precision
measurements of the properties of the Standard Model particles as well as rare processes rates.

Although the gauge sector of the Standard Model has been tested with high precision, the Higgs sector properties
are still greatly unknown. The signal strength of different production and decay channels are in overall agreement
with the Standard Model, but the errors are still large, and the coupling of the Higgs with third generation quarks
and leptons is subject to big uncertainties. Indeed, while the central value of the production rate of the Higgs in
association with top quarks is currently somewhat larger than the SM value~\cite{Khachatryan:2016vau},\cite{Aaboud:2017jvq},\cite{CMS:2017vru},  
the central value of the production rate of the Higgs decaying into bottom quarks and produced in association with heavy gauge bosons, which seemed
to be low at run I,  is
now in rough agreement with the SM prediction~\cite{Khachatryan:2016vau},\cite{Aaboud:2017xsd},\cite{Sirunyan:2017elk}.  Since the errors in these
determinations are still quite large, it is interesting to consider the possibility
that the couplings of the Higgs to top and bottom quarks differ from the SM values due to new physics effects.

In this article, we shall consider the possibility that not only the magnitude but also the sign of the Higgs
coupling to bottom quarks differ from the Standard Model predictions. This is an intriguing possibility that
could be realized in the simplest two Higgs doublet extension of the Standard Model~\cite{Ferreira:2014naa}. 
Such region of parameters has been invoked recently also in
models that lead to large rates of lepton flavor violating decays of the Higgs bosons $h \rightarrow \tau \mu$~\cite{Aloni:2015wvn}
and on theories of flavor at the weak scale~\cite{Bauer:2015fxa}.
 In this article we study the possible realization of this scenario within the minimal supersymmetric extensions of the SM,
namely the MSSM~\cite{mssmhiggs},\cite{Gunion:1984yn},\cite{mssmhiggsreview1},\cite{mssmhiggsreview2} and the
NMSSM\cite{Ellwanger:2009dp}.

Low energy supersymmetry~\cite{Martin:1997ns} leads to the stability of the weak scale under the large radiative effects induced
by possible heavy particles, like the ones associated with an hypothetical Grand Unified Theory (GUT). It also
leads to the radiative breaking of the electroweak symmetry, induced by corrections associated with the
superpartners of the third generation quarks. The low energy theory contains at least two Higgs doublets
and therefore the coupling of the Higgs to bottom quarks may be affected by mixing between the different
CP-even Higgs bosons in the theory. 

Large negative variation of the bottom quark coupling to the SM-like Higgs boson may be obtained in the MSSM at large values of $t_\beta$, the
standard ratio of the Higgs vacuum expectation values, and values of the heaviest CP-even Higgs boson masses
not far above the weak scale~\cite{Ferreira:2014naa}. This region of parameter space, however, is strongly restricted by
searches for Higgs bosons decaying into $\tau$-lepton pairs~\cite{Aaboud:2017sjh},\cite{CMS:2017epy}, which makes the realization of this scenario
difficult. As we shall show in this article, the scenario is more easily realized in the NMSSM, although the
necessary values of the coupling $\lambda$ of the singlet  to the doublet Higgs superfields are larger than the ones
leading to perturbative consistency of the theory up to the GUT scale.

This article is organized as follows.  In Section \ref{Type II} we analyze the possibility of a wrong-sign Yukawa coupling within two Higgs doublet
models. In Section \ref{MSSM/NMSSM} we study the possible realization of this idea within the MSSM and the NMSSM. After pointing out the
difficulties of its realization in the MSSM, in Section \ref{Results} we present an analytical and numerical analysis of this question within the NMSSM.
In Sections \ref{LHC} and \ref{Heavy Higgs} we study the experimental probes of this scenario. We reserve Section \ref{Conclusions} for our conclusions.

%\noindent Comment on MSSM Higgs mass\cite{mssmhiggsradcorr}\cite{mssmhiggsupperbound}\cite{cpviolatingmssmhiggs}.\\

\section{Wrong sign Yukawa  in Type II Two Higgs Doublet Models} \label{Type II}

\noindent The tree-level couplings of the lightest Higgs boson to Gauge bosons and fermions in type II 2HDM Higgs  are listed as below,
\bea
& g_{h V V} = s_{\beta - \alpha}\,,
\label{hlvvtree}  \\
& g_{h t\bar t}  =  \frac{m_t}{v}\,\frac{c_\alpha}{s_\beta} \equiv \frac{m_t}{v}\,\bigl(\sbma+\cbma t_\beta^{-1}\bigr)\, ,
\label{hltttree} \\
& g_{h b\bar b} = -\frac{m_b}{v}\,\frac{s_\alpha}{c_\beta} \equiv  \frac{m_b}{v}\,\bigl(\sbma-\cbma t_\beta\bigr)\,,
\label{hlbbtree}
\eea
where $s_{\alpha} \ (s_\beta)  = \sin\alpha \ (\sin\beta)$, $c_\alpha \ (c_\beta) = \cos\alpha \ (\cos\beta)$, $t_\beta = \tan\beta$  and $s_{\beta-\alpha} \ (c_{\beta-\alpha}) =
\sin(\beta-\alpha) \ (c_{\beta-\alpha})$.
As we can see from Eq.~(\ref{hltttree}), for the gauge boson couplings to be SM-like, we need $s_{\beta - \alpha} \approx 1$.  In this case, for
moderate values of $t_\beta$,  the Higgs coupling  to top-quarks or other up type fermions becomes SM-like due the $t_\beta$ suppression
of the second term on the right hand side of Eq.~(\ref{hltttree}). However, for the Higgs to b-quark coupling, Eq.~(\ref{hlbbtree}), a wrong sign could arise without changing the Higgs decay width and branching ratio when  $g_{h b\bar b}/g_{h b\bar b}^{SM} \simeq -1$. This could be achieved with minor changes of the Higgs couplings to top-quarks and weak gauge bosons for sizable values of  $t_\beta$  and~\cite{Ferreira:2017bnx},~\cite{Das:2017mnu}
\begin{equation}
t_\beta \ c_{\beta-\alpha} \approx 2.
\label{eq:wrongsign}
\end{equation}
This is in contrast with the condition $t_\beta c_{\beta - \alpha} \simeq 0$ that ensures a SM-like coupling of the bottom-quark to the Higgs boson.

The scalar potential of the most general two-Higgs-doublet extension of the SM may be written as :
\bea
\label{eq:generalpotential}
V &=& m_{11}^2 \Phi_1^\dagger \Phi_1+m_{22}^2 \Phi_2^\dagger \Phi_2-m_{12}^2 (\Phi_1^\dagger \Phi_2 +{\rm h.c.}) +\tfrac12 \lambda_1 ( \Phi_1^\dagger \Phi_1)^2+\tfrac12 \lambda_2 ( \Phi_2^\dagger \Phi_2)^2 \nonumber \\
&& +\lambda_3 ( \Phi_1^\dagger \Phi_1)( \Phi_2^\dagger \Phi_2)+\lambda_4 ( \Phi_1^\dagger \Phi_2)( \Phi_2^\dagger \Phi_1) \nonumber \\
&&+ \left\{ \tfrac12 \lambda_5 ( \Phi_1^\dagger \Phi_2)^2 + [ \lambda_6 (\Phi^\dagger_1\Phi_1)+ \lambda_7 (\Phi^\dagger_2\Phi_2)]\Phi_1^\dagger\Phi_2 + {\rm h.c.} \right\}\ ,
\eea
After converting to the Higgs basis~\cite{higgsbasis},\cite{Branco:1999fs}, the Higgs potential above could be rewritten as:
\be \label{higgsbasispot}
\mathcal{V} \supset  \ldots+\half Z_1(H_1^\dagger H_1)^2+\ldots +\big[Z_5(H_1^\dagger H_2)^2+Z_6 (H_1^\dagger
H_1) H_1^\dagger H_2+{\rm
h.c.}\bigr]+\ldots\,,
\ee
where we have only retained those terms relevant for the following discussion and the new couplings $Z_i's$ are associated with previous $\lambda_i's$ by the following relations~\cite{Gunion:2002zf},\cite{Craig:2013hca},\cite{Carena:2014nza}
\bea
Z_1 & \equiv & \lambda_1\cosbiv+\lambda_2\sinbiv+\half(\lambda_3+\lambda_4+\lambda_5)s^2_{2\beta}
+2s_{2\beta}\bigl[\cosbii\lambda_6+\sinbii\lambda_7\bigr]\,,\label{zeeone}\\
Z_5 & \equiv & \tfrac{1}{4} s^2_{2\beta}\bigl[\lambda_1+\lambda_2-2(\lambda_3+\lambda_4+\lambda_5)\bigr]+\lambda_5-s_{2\beta}c_{2\beta}(\lambda_6-\lambda_7)\,,\label{zeefive}\\
Z_6 & \equiv & -\half s_{2\beta}\bigl[\lambda_1\cosbii-\lambda_2\sinbii-(\lambda_3+\lambda_4+\lambda_5)c_{2\beta}\bigr]+\cosb c_{3\beta}\lambda_6+\sinb s_{3\beta}\lambda_7\,,\label{zeesix}
\eea

The CP-even Higgs mixing angle in this basis is identified with $\beta-\alpha$.  Consequently, we have~\cite{Gunion:2002zf},\cite{Carena:2014nza}
\be
c_{\beta-\alpha}=\frac{-Z_6 v^2}{\sqrt{(m_H^2-m_h^2)(m_H^2-Z_1 v^2)}}\,.\label{cbmaf}
\ee
%This term should be small 
%and can be realized in both the decoupling limit, i.e. $m_H \gg v$ and the alignment limit, i.e. $Z_6\approx 0 $.

As stressed before, since the observed Higgs boson has SM-like properties, $s_{\beta -\alpha} \simeq 1$,  in order to fulfill the requirement to obtain
a negative sign of the bottom Yukawa coupling, Eq.~(\ref{eq:wrongsign}), sizable values of $t_\beta$ are required.  For large values
of $t_\beta$, $s_\beta \simeq 1$, $c_\beta \simeq 1/t_\beta$ and $s_ {2 \beta} \simeq 2/t_\beta$. Since $Z_{1} v^2 \simeq m_{h}^{2}$, the denominator becomes approximately $m_{H}^{2} - m_{h}^{2}$.
From the relation of $Z_6$ to the quartic couplings $\lambda_i$ we obtain that, ignoring subdominant terms in $1/t_\beta$, an inversion of the sign of the Yukawa coupling leads to the following condition
\begin{equation}
[(\lambda_3 + \lambda_4 + \lambda_5) - \lambda_2 + \lambda_7 t_\beta] v^2  \simeq 2 (m_H^2 - m_h^2).
\label{eq:condition1}
\end{equation}
or, equivalently,
\begin{equation}
[(\lambda_3 + \lambda_4 + \lambda_5)  + \lambda_7 t_\beta] v^2  \simeq 2 m_H^2 - m_h^2.
\label{eq:condition2}
\end{equation}
where we have used the fact that, at large values of $t_\beta$, $m_h^2 \simeq \lambda_2 v^2$.  Hence, considering perturbative
values of the quartic couplings $\lambda_i$, it is straightforward to see that, unless $\lambda_7 \simeq {\cal{O}}(1)$, the values of
$m_H$ must be of order of a few hundred GeV.  

The relation between the charged and neutral Higgs masses is given by
\begin{equation}
m_{H^{\pm}}^2 = m_A^2 + \frac{v^2}{2} ( \lambda_5 - \lambda_4)
\label{eq:CPoddCh}
\end{equation}
and hence large values of $\lambda_4$ and $\lambda_5$ may induce a large splitting between the charged and the CP-odd Higgs boson masses.

\section{Wrong sign Yukawa couplings in the MSSM and the NMSSM} \label{MSSM/NMSSM}

\subsection{MSSM and minimal NMSSM}

The tree-level Higgs sector of the MSSM is a type-II 2HDM and consists of two Higgs doublets with quartic couplings which are related to the squares of the weak gauge couplings. Since Supersymmetry imposes concrete values for  the quartic couplings $\lambda_i$  it is interesting to check whether the wrong sign Yukawa coupling could arise in the frame of the MSSM without conflicting with other Higgs phenomenology. For this, one has to take into account the relevant radiative corrections arising from the interaction of the Higgs field with the third generation fermions and their scalar superpartners.  In the MSSM, it's usually argued that a SM-like  neutral Higgs boson could only be obtained in two distinct scenarios, i.e. the decoupling limit~\cite{Gunion:2002zf}-\cite{Haber:2013mia} and the alignment limit~\cite{Gunion:2002zf},\cite{Craig:2013hca},\cite{Haber:2013mia},\cite{Carena:2001bg}. The decoupling limit happens when $m_h \ll m_H $, while the alignment limit arises when one of the CP-even Higgs bosons, when expressed as a linear combination of the real parts of the two neutral Higgs fields, lies in the same direction in the two Higgs doublet field space as  neutral Higgs vacuum expectation values. This alignment does not in general depend on the masses of the non-standard Higgs bosons. 
The region of parameters under investigation requires a nonvanishing value of $c_{\beta - \alpha}$ and therefore a departure from the alignment
limit.  Hence, some departures from the SM behavior of the lightest Higgs are expected.% beyond the ones associated with the change of the bottom coupling. 
%However, as stressed before, for sizable values of $t_\beta$, the condition $t_\beta c_{\beta-\alpha} = 2$, Eq.~(\ref{eq:wrongsign}), can be fulfilled for $s_{\beta-\alpha} \simeq 1$.

In the MSSM,  it's not difficult to work out an approximate expression for $Z_6$ at the one-loop level. Taking into account that the most relevant 
radiative corrections may be absorbed in the definition of the Higgs mass at large values of $t_\beta$, one gets
\begin{eqnarray}
\lambda_2  & \simeq  & \frac{m_h^2}{v^2}
\nonumber\\
\lambda_3 & \simeq & \frac{1}{4}\left( g_2^2 - g_1^2 \right)
\nonumber\\
\lambda_4 & \simeq &  - \frac{1}{2} g_2^2
\nonumber\\
\lambda_7 & \simeq & \frac{3h_t^4 }{16\pi^2 
}\frac{\mu}{M_S}\left(\frac{A_t^3}{6M_S^3}-\frac{A_t}{M_S}\right)
\nonumber\\
m_t & \simeq & h_t \frac{v}{\sqrt{2}}
\end{eqnarray}
~\\
%{\bf Add some references}
%~\\
%The largest one-loop contributions are proportional to the fourth power of the top-quark Yukawa coupling $h_t$, namely:
%\be
%Z_6 v^2 = -s_{2\beta}\left\{m_Z^2 c_{2\beta}-\frac{3v^2 s_\beta^2  h_t^4}{16\pi^2}\biggl[\ln\left(\frac{M_S^2}{m_t^2}\right)+\frac{X_t(X_t+Y_t)}{2M_S^2}-\frac{X_t^3 Y_t}{12 M_S^4}\biggr]\right\},
%\label{zeesixcorr}
%\ee
where $m_t$ is the top quark mass, $M_S$ is the stop mass scale, $A_t$ is the trilinear stop-Higgs coupling and $\mu$ is the Higgsino mass parameter
(for a more complete expression, see Refs.~\cite{Haber:1993an},~\cite{Carena:1995bx},~\cite{Lee:2015uza}).
Taking into account Eq.~(\ref{cbmaf}), we  get  the following estimate~\cite{Carena:2014nza}:
\be
t_\beta\; c_{\beta-\alpha}\simeq \frac{1}{m_H^2-m_h^2}\left[-m_h^2-m_Z^2+
\frac{3m_t^4 }{4\pi^2 v^2
}\frac{\mu}{M_S}\left(\frac{A_t^3}{6M_S^3}-\frac{A_t}{M_S}\right) t_\beta \right] \,
\label{tbcba_mssm}
\ee
where the first two terms inside the
square bracket comes from  $(\lambda_3 + \lambda_4 + \lambda_5 - \lambda_2) v^2$, while the last term comes from
the radiatively induced $\lambda_7 v^2 t_\beta$ contribution.  

If we want $t_\beta\ c_{\beta-\alpha}$ to be as large as 2, it's clear that we need the third term in the square bracket to be quite large. Unfortunately, this will lead to an unacceptably large value of $t_\beta$, which pushes the Yukawa coupling to third generation down-type
fermions to large values that are restricted by heavy Higgs searches~\cite{Aaboud:2017sjh},\cite{CMS:2017epy}. 
In order to see that, let's recall the fact that stability of the Higgs potential demands that $|A_t|$ and $|\mu|$ should both be smaller than 3~$M_S$~\cite{Blinov:2013fta}. Under this constraint, the maximum of the expression $\frac{\mu}{M_S}\left(\frac{A_t^3}{6M_S^3}-\frac{A_t}{M_S}\right)$ is  $4.5$, which
is achieved for  $A_t/M_S=3$ and $\mu/M_S = 3$. When normalized in terms of the square of the Higgs mass,  the coefficient $\frac{3m_t^4 }{4\pi^2 v^2}$ is about $m_h^2/16$, which is very  small compared to the first two positive terms in the square bracket. Thus, for $t_\beta\ c_{\beta-\alpha}$ to reach the target value $2$, $t_\beta$ needs to be very large. More specifically, for $m_H \approx 250~GeV$, one requires values of $t_\beta \approx 30$, while for $m_H \approx 500~GeV$,  $t_\beta \approx 120$.  These values
of $m_H$ and $t_\beta$ are excluded by heavy Higgs searches at the LHC~\cite{Aaboud:2017sjh},\cite{CMS:2017epy}.  One could avoid these constraints for larger values of
the heavy Higgs mass, larger than 1~TeV. However, for $m_H \approx 1$~TeV one requires $t_\beta \approx 500$, and it is difficult to
keep the perturbative consistency of the theory at such large values of $t_\beta$.

In order to address the question of perturbative consistency of the MSSM at large values of $t_\beta$, we should stress that in supersymmetric models there are relevant radiatively generated couplings of
the $H_u$ Higgs boson to the bottom quarks, which imply a departure from the simple type II 2HDM. These modifications are particularly important for large values of the Higgsino
mass parameter $\mu$ and lead to a modification of $\kappa_b$~\cite{Carena:1998gk}:
  
\begin{equation}
\kappa_b = \frac{g_{hbb}}{g_{hbb}^{\rm SM}} = - \frac{s_\alpha}{c_\beta} \left[ 1 - \frac{\Delta_b}{1 + \Delta_b} \left( 1 + \frac{1}{t_\alpha t_\beta}\right)\right]
\label{eq:kappab}
\end{equation}
where $\Delta_b$ is given by~\cite{Hempfling:1993kv},\cite{Hall:1993gn},\cite{Carena:1994bv}
\begin{equation}
\Delta_b \simeq  \left( \frac{2 \alpha_3}{3 \pi} M_3 \mu \ I(m_{\tilde{b}_1},m_{\tilde{b}_2},M_3)  + \frac{h_t^2}{(4 \pi)^2} A_t \mu \  I(m_{\tilde{t}_1}, m_{\tilde{t}_2},\mu) \right) t_\beta
\end{equation}
and the function $I(a,b,c)$ is given by%~\cite{Hall:1993gn} % {\bf Is the I(a,b,c) equation correct? in [37] looks like a,b,c instead of $a^{2},b^{2},c^{2}$.}
\begin{equation}
I(a,b,c) = \frac{ a^2 b^2 \ln(a^2/b^2) + b^2 c^2 \ln(b^2/c^2)+c^2a^2\ln(c^2/a^2)}{(a^2-c^2)(a^2-b^2)(b^2-c^2)}
\end{equation} 
and $M_3$, $m_{\tilde{b}_i}$, $m_{\tilde{t}_i}$ are the gluino, sbottom and stop mass eigenvalues.  

There are similar corrections to the tau coupling, but they are governed by weak coupling effects and are therefore  less significant. The above
corrections imply a difference between $\kappa_b$ and $\kappa_\tau$ and therefore have relevant phenomenological consequences for 
sizable values of $t_\beta$.  In particular, in the region of parameters under investigation, $\kappa_\tau$ tends to differ from 
$\kappa_b$ by a few tens of percent. 

Moreover,  the coupling of the heavy MSSM-like Higgs bosons to bottom quarks becomes
\begin{equation}
g_{Hb\bar{b}} \simeq g_{Ab\bar{b}} \simeq h_b \simeq \frac{m_b  t_\beta}{v (1 + \Delta_b)}.
\label{eq:hb}
\end{equation}
These corrections must be in general considered when studying the production and decay of the heavy CP-even and CP-odd Higgs bosons and lead to some
moderate modification of these rates with respect to the ones expected in the type II 2HDM.  

At very large values of $t_\beta$, the bottom Yukawa coupling $h_b$ and the $\tau$ Yukawa coupling 
$h_\tau \simeq m_\tau t_\beta/v$ become large.  For positive values of $\Delta_b$, the increase of the bottom-Yukawa coupling is 
slower than what is expected using the tree-level relations and hence perturbative consistency can be kept for larger values of $t_\beta$.
However, for sizable $\Delta_b$, it can be easily shown that the condition to invert the sign of the bottom Yukawa coupling becomes
\begin{equation}
t_\beta c_{\beta-\alpha}  = 2 ( 1 + \Delta_b).
\label{eq:modcond}
\end{equation}
Therefore, for very large values of $t_\beta$ there is a tension between maintaining the perturbative consistency of the theory, which depends on $h_b$  
and as shown in Eq.~(\ref{eq:hb}) is more easily fulfilled for  positive $\Delta_b$,   and the fulfillment of Eqs.~(\ref{tbcba_mssm})
and~(\ref{eq:modcond}).  
Thus we reach the conclusion that within the MSSM the current LHC bounds make it very difficult to invert the sign of the Higgs coupling to bottom quarks 
while keeping the perturbative consistency of the theory at low energies.

Next let's turn to the Next-to-Minimal supersymmetric extension of the SM~(NMSSM)~\cite{Ellwanger:2009dp}, with only an extra singlet superfield added on top of the MSSM. The CP-even singlet will mix with the two neutral CP-even Higgs bosons. We consider first the simpler case when the superpotential is scale invariant and  thus the complete Lagrangian would have an accidental $\mathbb{Z}_3$ symmetry. The superpotential is given by
\begin{equation}
W = \lambda \widehat{S} \widehat{H}_u\cdot \widehat{H}_d + \frac{\kappa}{3} \widehat{S}^{\,3}  + h_u \widehat{Q}\cdot \widehat{H}_u\, \widehat{U}_R^c + h_d \widehat{H}_d\cdot\widehat{Q}\, \widehat{D}_R^c + h_\ell \widehat{H}_d \cdot\widehat{L}\,\widehat{E}_R^c\,,
\end{equation}
where $\widehat{S}, \widehat{H_u}, \widehat{H_d}$ denote the singlet and doublet Higgs superfields, and $\widehat{Q},\widehat{D}_R,\widehat{U}_R$ are the quark superfields,
while $\widehat{L},\widehat{E}_R$ are the lepton superfields, $h_i$ are the Yukawa couplings and
$\lambda$ and $\kappa$ are both dimensionless couplings. Note that in this case, $\mu$ is an effective mass parameter generated by the vev of the singlet, $\mu_{\text{eff}} = \lambda s$; we will use $\mu$ in the following discussions to refer to $\mu_{\text{eff}}$.  Observe that the fields $H_u$ and $H_d$ have opposite hypercharge. These
fields can be related to the fields $\Phi_1$ and $\Phi_2$ introduced before by the relations
\begin{equation}
H_d^i = \epsilon_{ij} \Phi_1^{j*} \;\;\;\;\;\;\;\;\;\;  H_u^i = \Phi_2^i
\end{equation} 
and therefore
\begin{equation}
H_u H_d  = -\Phi_2^\dagger \Phi_1
\end{equation}

The most significant change in the NMSSM would be that at tree level, 
\begin{equation}
\delta \lambda_4 = \lambda^2
\label{eq:deltalambda4}
\end{equation} 
and therefore there is an extra correction proportional to $\frac{1}{2} \lambda^2 v^2 s_{2\beta}^2$  in the $M_{11}^2$ term of the Higgs basis. This term is relevant since it can lift up the upper limit of the lightest Higgs mass at tree level~\cite{Espinosa:1991gr}, thus making it possible to reach $125~$GeV without the large quantum corrections needed in the MSSM~\cite{mssmhiggsradcorr},\cite{mssmhiggsupperbound},\cite{Bagnaschi:2014rsa},\cite{Hahn:2013ria}--\cite{Bahl:2016brp}.
What's more important in this case is that it can modify the $Z_6$ term introduced earlier in the MSSM case and release the strong tension between $t_\beta$ and $M_A$ present  in the MSSM to make $c_{\beta-\alpha}t_\beta=2$ feasible. In the NMSSM, 
considering heavy singlets,  it's straightforward to get the expression for $Z_6$ at moderate or large values of $t_\beta$, including only the stop loop corrections, namely~\cite{Carena:2015moc}:
\be
Z_6 v^2 \approx \frac{1}{t_\beta} \left[m_h^2 + m_Z^2-\lambda ^2 v^2 \right] +\frac{3 v^2  h_t^4 \mu X_t}{16 \pi ^2 M_S^2} \left(1-\frac{X_t^2}{6M_S^2}\right)
\ee
where $X_t = A_t - \mu/t_\beta$, which leads to
\be
t_\beta\; c_{\beta-\alpha} \approx \frac{-1}{m_H^2-m_h^2}\left[\left(m_h^2 + m_Z^2-\lambda ^2 v^2\right) +\frac{3 m_t^4 A_t \mu t_\beta}{4 \pi ^2 v^2 M_S^2} \left(1-\frac{A_t^2}{6M_S^2}\right)\right]
\label{tbcba_NMSSM}
\ee
Compared with Equation (\ref{tbcba_mssm}), we have an extra $-\lambda ^2 v^2$ term in the parenthesis,  which comes from $\delta \lambda_4$, Eq.~(\ref{eq:deltalambda4}),
and tends to push $t_\beta\; c_{\beta-\alpha}$ towards positive values, making it promising to get $t_\beta\; c_{\beta-\alpha} = 2$ with smaller values of $t_\beta$. However, for that purpose we need $\lambda$ to be of order 1. For $\lambda$ or $\kappa$ of order 1, one can no longer neglect 
the chargino, neutralino and Higgs loop contributions when evaluating the Higgs mass and couplings. Although approximate analytical expressions exist in this case, the formulae become complicated and beyond some approximate formulae we will present in the next section, we will mostly base our results in a numerical analysis with full quantum corrections up to two-loop level, which are necessary to select the proper region of parameter space leading to the inversion of the bottom coupling. Moreover, large $\lambda$ could lead to a Landau pole problem at energies lower than the Grand Unification scale.  We reserve a discussion of this issue to the Appendix. 

\subsection{Moderate values of $t_\beta$}
\label{sec:Moderate}

As we discussed before, the MSSM is highly constrained for the large values of $t_\beta$ necessary to achieve an inversion of the bottom couplings.
These constraints tend to translate into the NMSSM since for very large values of $t_\beta$ the decay of the MSSM-like CP-even and CP-odd Higgs bosons
into bottom-quarks and $\tau$-leptons remain relevant.  In the NMSSM, however, for sufficiently large values of $\lambda$, an inversion of the sign of 
the SM-like Higgs boson coupling to bottom quarks may be achieved at moderate values of $t_\beta \simeq 5$--$10$.  For this range of values of $t_\beta$
the constraints from direct searches for heavy Higgs bosons become weaker, but some of the approximations performed before cease to be valid. In particular,
the effects of the mixing between the CP-even Higgs bosons become relevant and cannot be dismissed. Due to these mixing effects, and ignoring the mixing
with singlets, we obtain
\begin{equation}
m_h^2 \simeq Z_1 v^2 - \frac{Z_6^2 v^4}{m_H^2}
\end{equation}
or, equivalently
\begin{equation}
m_h^2 \simeq Z_1 v^2 - c_{\beta -\alpha}^2 m_H^2 \simeq \lambda_2 v^2 - \frac{4 m_H^2}{t_\beta^2}.
\end{equation}
For very large values of $t_\beta$ the last term may be safely ignored. However, for moderate values of $t_\beta$ this term cannot
be ignored and tends to push the mass of the Higgs boson to values that are below the experimentally observed value. Mixing with the
singlets only worsen this situation. In order to address this problem, a departure from the simple $\mathbb{Z}_3$ invariant NMSSM is necessary.

\subsection{NMSSM with singlet tadpole terms} \label{NMSSM_singlet}

% As we will show in later sections, beyond the problems associated with perturbativity,
 As discussed above, we are interested in the inversion of the sign of the coupling of the bottom-quark to the
 Higgs in the simplest NMSSM case, with sizable values of $\lambda$ and moderate values 
 of $t_\beta$.   This simple framework tends to lead to problems in the
CP-even Higgs sector, since  as we discussed in the previous section, the lightest CP-even Higgs boson mass 
is generically pushed to values below the experimentally observed ones due to large mixing
effects.   A possible solution to this problem is to include a non-zero singlet tadpole term $\xi_S$ to the potential
\begin{equation}
\Delta V = \ \xi_S \ S + h.c.
\end{equation}
This  term, which could be a result of the supersymmetry breaking mechanism at high scales~\cite{Panagiotakopoulos:1998yw},\cite{Panagiotakopoulos:1999ah},\cite{Panagiotakopoulos:2000wp},  serves 
to break the $\mathbb{Z}_3$ symmetry explicitly and get rid of unwanted  domain walls.  
For values of $\mu$ of the order of the weak scale and $\lambda$ couplings of order one, a sizable value of $|\xi_S|$ leads to large values of
the singlet mass. Ignoring other terms that become subdominant for large value of $|\xi_S|$, one obtains
\begin{equation}
\left< S \right> = \frac{\mu}{\lambda} \simeq -\frac{\xi_S}{m_S^2}
\end{equation}
or, equivalently
\begin{equation}
m_S^2 \simeq - \frac{\lambda \xi_S }{\mu}
\label{mS2}
\end{equation}
A sizable  $|\xi_S|$ could keep the singlet decoupled from the two neutral Higgs bosons, reducing the problem to an approximate 2x2 Higgs mixing one, with low energy quartic couplings that are modified by terms proportional to powers of the couplings $\lambda$ and $\kappa$.  For moderate
values of $\xi_S$,  the decoupling effects may affect the low energy theory in a relevant way. We shall discuss these effects in more detail below. 

An additional consequence of large values of $\xi_{S}$ is that the singlet mass may become much larger than the mass of the singlino. In this case, the quartic coupling of $H_{u}$ has sizable corrections produced by $\lambda^{4}$ loop contributions from singlets and singlinos. The correction to $\lambda_{2}$ from these contributions is given by
\begin{equation}
\delta \lambda_{2} \simeq \frac{\lambda^{4}}{16 \pi^{2}} \ln \Big( \frac{m_{S}^{2}}{\mu^{2}} \Big)  \simeq \frac{\lambda^{4}}{16 \pi^{2}} \ln \left( \left|\frac{\lambda \xi_S}{\mu^{3}} \right|\right)
\label{dlam2}
\end{equation}
where we have used  the expression for $m_{S}^{2}$ given in Eq.~(\ref{mS2}).  It is therefore clear that for values of $|\mu|$ of the order of the weak scale, large values of $\xi_{S}$ result in large positive corrections to $\lambda_{2}$.  
These corrections  can compensate the negative contributions to the Higgs mass induced by mixing effects and 
constrain the allowable values of $\xi_{S}$ via the experimental constraints on the lightest CP-even Higgs mass, which will be examined in more specificity in the next section. 

A further possible modification to the NMSSM is the inclusion of a similar tadpole term in the superpotential, namely beyond the trilinear terms associated with the Yukawa, $\lambda$ and $\kappa$ couplings,
one may add a tadpole term of the form~\cite{Ellwanger:2009dp}
\begin{equation}
\delta W =  \xi_F S
\end{equation}
where $\xi_F$ is a dimension 2 parameter. One action of such a term, as we shall discuss,  is to modify the spectral relationships between 
the neutral and charged Higgs bosons. In our initial analysis, we first set $\xi_{F}=0$; however,  we shall discuss the impact of this term  in later examinations of 
pseudoscalar decays in Section \ref{Heavy Higgs}.

The decoupling of the singlet induces corrections to $\lambda_{4}$ and $\lambda_{5}$, and a sizable correction to the quartic coupling $\lambda_7$.  This can be seen by ignoring subdominant terms and reducing the singlet-dependent terms in the scalar potential to
\begin{equation}
( m_S^2 + \lambda^2 | H_u|^2 )  |S|^2  + [ S ( \lambda A_\lambda H_u H_d + \xi_S) + h.c. ] + \left| \xi_F +\lambda H_u H_d + \kappa S^2 \right|^2
\label{eq:Lag}
\end{equation}
where we shall assume that, due to the effect of the tadpole terms, $m_S^2$ is much larger than $\lambda^2 H_u^2 \simeq \lambda^2 v^2$. From Eq.~(\ref{eq:Lag}), and ignoring small corrections induced by the vacuum expectation
values of the singlet and doublet fields,  we can see that the masses of the 
CP-even and CP-odd singlet eigenstates are approximately given by
\begin{equation}
m_{h_S}^2 = m_S^2 + 2 \xi_F \kappa, \;\;\;\;\;\;\;\;\;\;\;\;  m_{A_S}^2 = m_S^2 - 2 \xi_F \kappa, 
\end{equation} 
repsectively. Now, one can integrate out the singlets, replacing the singlet fields by their equation of motion. This is roughly 
\begin{equation}
{\rm Re}(S) \simeq  -\frac{\lambda A_\lambda (H_u H_d + h.c.) + \xi_S }{2 \ m_{h_S}^2}, \;\;\;\;\;\;\;\;\;\;\;
{\rm Im}(S) \simeq  -i\frac{\lambda A_\lambda (H_u H_d-h.c.)   }{2 \ m_{A_S}^2} .
\end{equation}
Replacing this expression into the orginal Lagrangian density, Eq.~(\ref{eq:Lag}),  one obtains contributions to $\lambda_4$, $\lambda_5$, and $\lambda_7$ given by
\begin{eqnarray}
\delta \lambda_4  & \simeq & -  \lambda^2 \left(\frac{A_\lambda^2}{2 m_{h_S}^2} + \frac{A_\lambda^2}{2 m_{A_S}^2} \right) + 2 \lambda^2 \kappa \frac{\xi_S A_\lambda}{m_{h_S}^2} \left( \frac{1}{m_{h_S}^2} - \frac{1}{m_{A_S}^2}\right) 
\nonumber\\
&+& \frac{\xi_F A_\lambda^2 \kappa \lambda^2}{2}\left( \frac{1}{m_{h_S}^4} - \frac{1}{m_{A_S}^4}\right) 
+ \frac{\kappa^2 \lambda^2 A_\lambda^2 \xi_S^2}{m_{h_S}^4}\left( \frac{3}{m_{h_S}^4} + \frac{1}{m_{A_S}^4} \right)
\nonumber\\
\delta \lambda_5  & \simeq & -  \lambda^2 \left(\frac{A_\lambda^2}{2 m_{h_S}^2} - \frac{A_\lambda^2}{2 m_{A_S}^2} \right) + 2 \lambda^2 \kappa \frac{\xi_S A_\lambda}{m_{h_S}^2} \left( \frac{1}{m_{h_S}^2} + \frac{1}{m_{A_S}^2}\right) 
\nonumber\\
&+& \frac{\xi_F A_\lambda^2 \kappa \lambda^2}{2}\left( \frac{1}{m_{h_S}^4} + \frac{1}{m_{A_S}^4}\right) 
+ \frac{\kappa^2 \lambda^2 A_\lambda^2 \xi_S^2}{m_{h_S}^4}\left( \frac{3}{m_{h_S}^4} - \frac{1}{m_{A_S}^4} \right)
\nonumber\\
\delta \lambda_7 & \simeq & -\lambda^3 \frac{\xi_S A_\lambda}{m_{h_S}^4}  
\label{eq:singdec}
\end{eqnarray}

The value of $\mu A_\lambda$ is related to the CP-odd Higgs spectrum by
the relation
\begin{equation}
\left[\mu \left(A_\lambda + \frac{\kappa}{\lambda} \mu \right) + \lambda \xi_F \right] t_\beta \simeq M_A^2
\label{eq:MA_xif}
\end{equation}
and therefore for fixed $M_{A}$ and moderate values of $\kappa$, sizable negative values of $\xi_{F}$ result in large positive values of $\mu A_{\lambda}$. Hence,  for values of $\mu$ at the weak scale, the presence of negative $\xi_{F}$ can lead to sizable  values of $A_\lambda$ and therefore to large corrections to $\lambda_{7}$. Such large
corrections may induce a modification of the value of $c_{\beta - \alpha}$, which including only the dominant terms becomes
\begin{equation}
\begin{split}
t_\beta\; c_{\beta-\alpha} \approx \frac{1}{m_H^2-m_h^2} \Bigg[\left(\lambda ^2 \left( 1 - \frac{A_\lambda^2}{m_{h_S}^2} \right) v^2- \lambda_2 v^2 - M_Z^2\right) & +\frac{3 m_t^4 A_t \mu t_\beta}{4 \pi ^2 v^2 M_S^2} \left(\frac{A_t^2}{6M_S^2}-1\right) \\
& - \lambda^3 v^2 \frac{\xi_S A_\lambda t_\beta}{m_{h_S}^4}  \Bigg].
\end{split}
\label{eq:tbcba_singlet}
\end{equation}
Hence, the reduction of the $\lambda^2$ contribution due to sizable values of $A_\lambda$ may be compensated by the explicit $\xi_S$ dependence appearing in the last
term of Eq.~(\ref{eq:tbcba_singlet}).

Moreover, including the above corrections to $\lambda_{4}$ and $\lambda_{5}$ modifies the difference between the charged and neutral CP-odd Higgs
boson masses, Eq.~(\ref{eq:CPoddCh}),
\begin{equation}
( \lambda_5 - \lambda_4) \frac{v^2}{2} \simeq M_W^2 + \left\{\left(\frac{A_\lambda^2}{m_{A_S}^2} -1 \right) +   \kappa \left[\frac{4 \xi_S A_\lambda}{m_{A_S}^2m_{h_S}^2} +
 \left(\xi_F - \frac{2\kappa \xi_S^2}{m_{h_S}^4}\right)  \frac{A_\lambda^2}{m_{A_S}^4}\right] \right\} \frac{\lambda^2 v^2}{2}.
\label{newsplit}
\end{equation}
Hence for sizable $A_\lambda$, the splitting between
the CP-odd and the charged Higgs mass induced by the large values of $\lambda$, Eq.~(\ref{eq:CPoddCh}), may be reduced by the effects associated
with the singlet decoupling. These observations will be important in examining the constraints from precision measurements and the decay mode $A_{1} \to hZ$.

\section{NMSSM Results: Full Analysis} \label{Results}

%As mentioned in the last section, besides the large one-loop quantum corrections to $\mathcal{M}^2_{11}$, there are relevant two-loop corrections in the NMSSM from various sources, which also give non-trivial contributions to the Higgs mass matrix at large values of $\lambda$, $\kappa$ and that modify the resulting bottom coupling to the SM-like Higgs in a significant way,  by adjusting the mixing between the two light CP-even Higgs bosons. From the full expression of the squared mass matrix elements at  two-loop level~\cite{Ellwanger:2009dp}, it's straightforward to see that the most important $\kappa$ or $\lambda$ dependent contributions come from Higgs and chargino/neutralino loops. After a short numerical check, we found that for values of $\kappa$ and $\lambda$ of order one, these Higgs and electroweakino-loop corrections could be as large as a few tens of percent of the unperturbed squared mass matrix elements. This observation points to the fact that, as said before,  in order to find the accurate Higgs mixing and Yukawa couplings at large values of $\kappa$ and $\lambda$, one has to take these loop corrections into account.
%\textbf{ I think this paragraph could be trimmed down a bit }

In this section, we conducted a numerical search for a wrong-sign Yukawa coupling within the NMSSM model using the NMSSMTools code~\cite{Ellwanger:2004xm}, which includes the most relevant one and  two-loop radiative corrections to the Higgs mass matrix elements. In this calculation, we scanned over 6 independent parameters: $t_\beta, M_A, \mu, \lambda$, $\kappa$, and $\xi_{S}$. We fixed the gaugino and third generation scalar mass parameters to values that are not constrained by present experimental bounds and that contribute to the obtention of a proper SM-like Higgs mass at sizable
values of $t_\beta$.  This selection is somewhat arbitrary and does not play a relevant role in the Higgs phenomenology at moderate values of $t_\beta$.
However, as we will discuss below, it has an impact on the analysis of the flavor and Dark Matter constraints.  In our analysis 
we set $A_t = 1700$ GeV, $A_\tau$,$A_b=1500$~GeV, and the squark and slepton masses at $M_S = 1$~TeV. The weak gauginos were assumed to be
heavy, $M_{1,2} =1$~TeV, while the gluino mass was fixed at $M_3 = 2$~TeV.  

As discussed in the previous section, taking into account the
strong  constraints on large $t_\beta$ for the relatively low values of the MSSM-like CP-odd Higgs mass~\cite{Aaboud:2017sjh},\cite{CMS:2017epy}
necessary to induce a large correction to the bottom coupling, we  concentrate on moderate values of $t_\beta$. In particular,  values of $t_\beta$  in the range 6--10 and
values of the CP-odd Higgs mass $m_A$  within the interval of $300$~GeV to $500$~GeV are preferred due to these considerations. While larger values of $m_A$ 
make it more difficult to obtain a large correction to the bottom coupling, smaller values of $m_A$ lead to tension with the current CP-even and CP-odd neutral Higgs
searches and, for $\xi_F = 0$, to low values of the charged Higgs mass, excluded by top-quark decay studies. 

\begin{figure}[H]
    \centering
    \includegraphics[width=10cm]{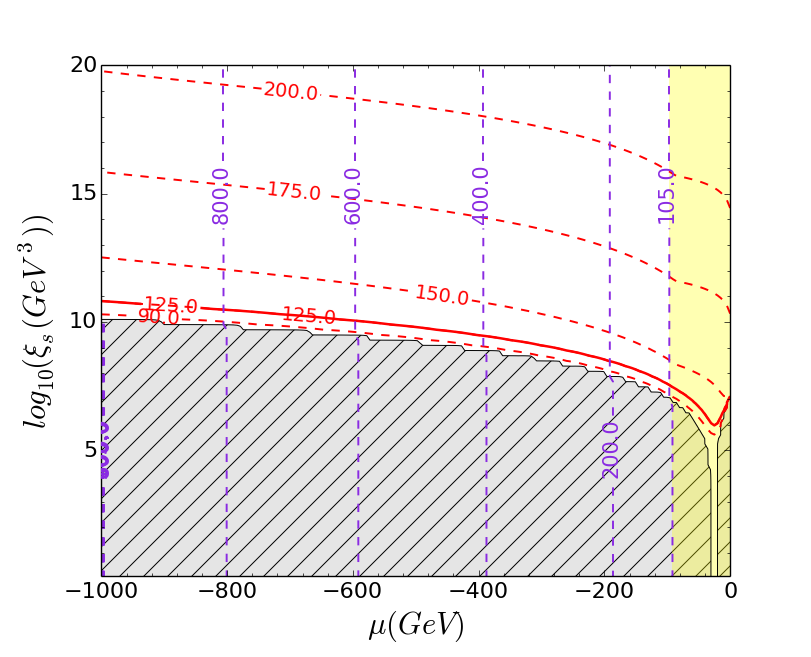}
    \caption{In this plot, $\lambda=1.3$ and $\kappa=0.1$ are fixed, and $\mu$ and $\xi_s$ are varied to examine the allowable values of these parameters. The gray area is excluded for negative Higgs mass. Red contour lines indicate values of the lightest Higgs mass, with 125 GeV represented by the solid contour. The lightest chargino mass contours are displayed in purple.}
    \label{NMSSM_tadpole}
\end{figure}

%Because we wish to decouple the singlet from the Higgs doublets in order to gain an effective 2HDM, we require sizable values of $\xi_{S}$. The $\lambda^{4}$ corrections to $\lambda_{2}$ from singlet- and singlino-mediated loops, Eq.~(\ref{dlam2}), become relevant, and in particular provide a positive contribution to the value of the lightest CP-even Higgs mass.  
In order to identify the possible ranges of $\xi_{S}$, we first performed a scan over a wide range of values of $\xi_{S}$ and $\mu$ for 
some characteristic parameters of the theory leading to a variation of the coupling of the Higgs to the bottom-quark.  An examination of Eq.~(\ref{eq:tbcba_singlet}) 
indicates that to obtain $t_\beta c_{\beta - \alpha} = 2$, one requires negative values of $\mu \times \xi_S$. The values of $|\mu|$ are required to be at the weak scale and lead to allowed values of the chargino and neutralino masses.  Taking $\xi_S$ to be positive,
we therefore scan a range for $\mu$ from -1000 GeV to 0 GeV. All other parameters are fixed at values that favor positive values of $t_\beta c_{\beta - \alpha}$: $\lambda = 1.3$, $\kappa = 0.1$, $M_{A} = 350$, and $t_\beta = 7$.  The result of this scan is shown in Fig. 1. We find that the value of  the lightest CP-even Higgs mass $m_{1}$ does indeed increase with $\xi_{S}$, as expected from Eq.~(\ref{dlam2}). The contour with $m_{1}=125$ GeV is indicated by the solid red line; the constraint of $m_{1}=125 \pm 3$ GeV, which takes into account  the uncertainty in the determination of the Higgs mass,  therefore constrains the value of $\xi_{S}$ to be on the order of $10^{9}$ to $10^{11}$ GeV$^{3}$ for values of $|\mu|$ at the weak scale.

\begin{figure}[H]
    \centering
    \includegraphics[width=.65\textwidth]{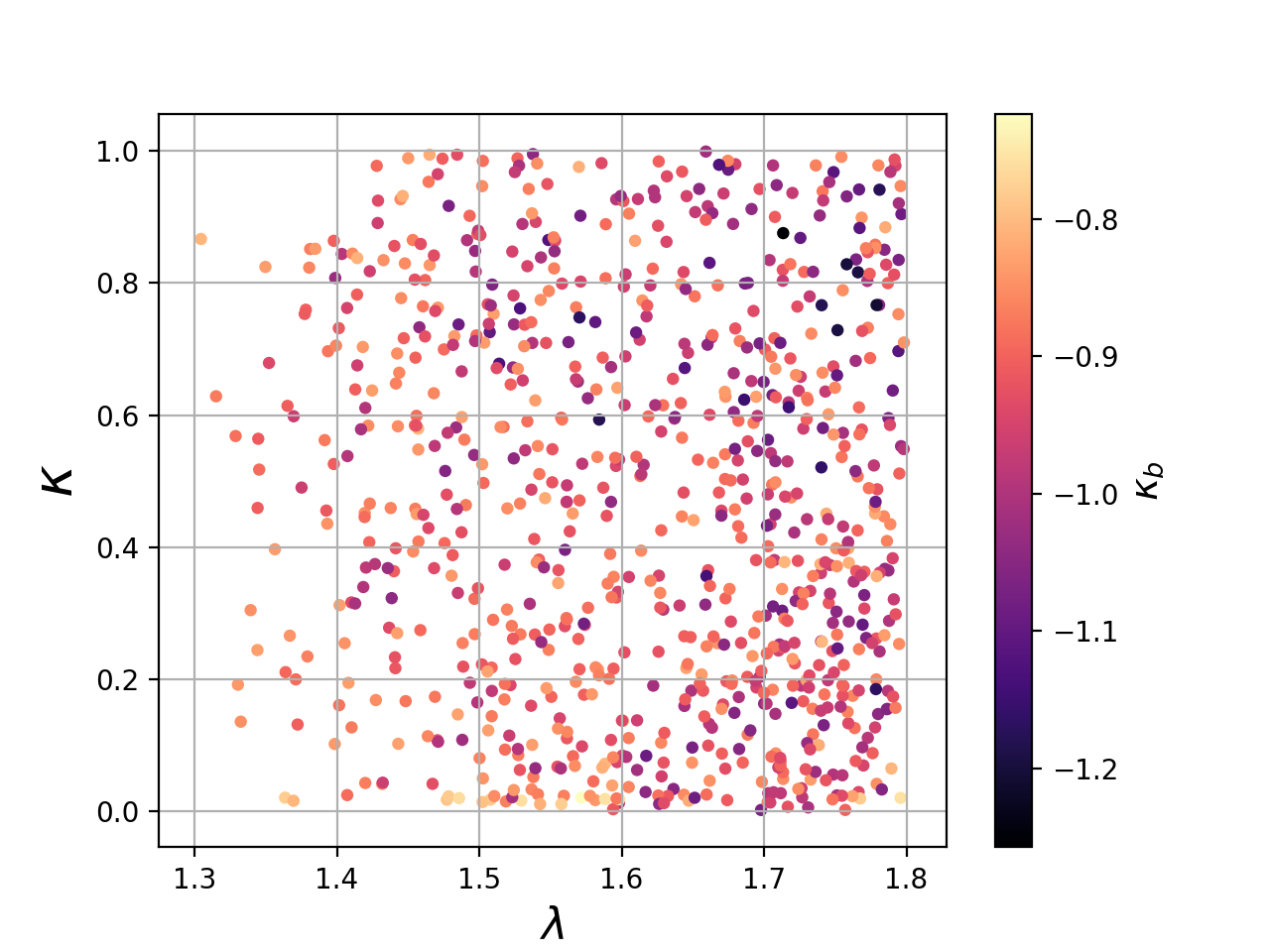}
    \includegraphics[width=.65\textwidth]{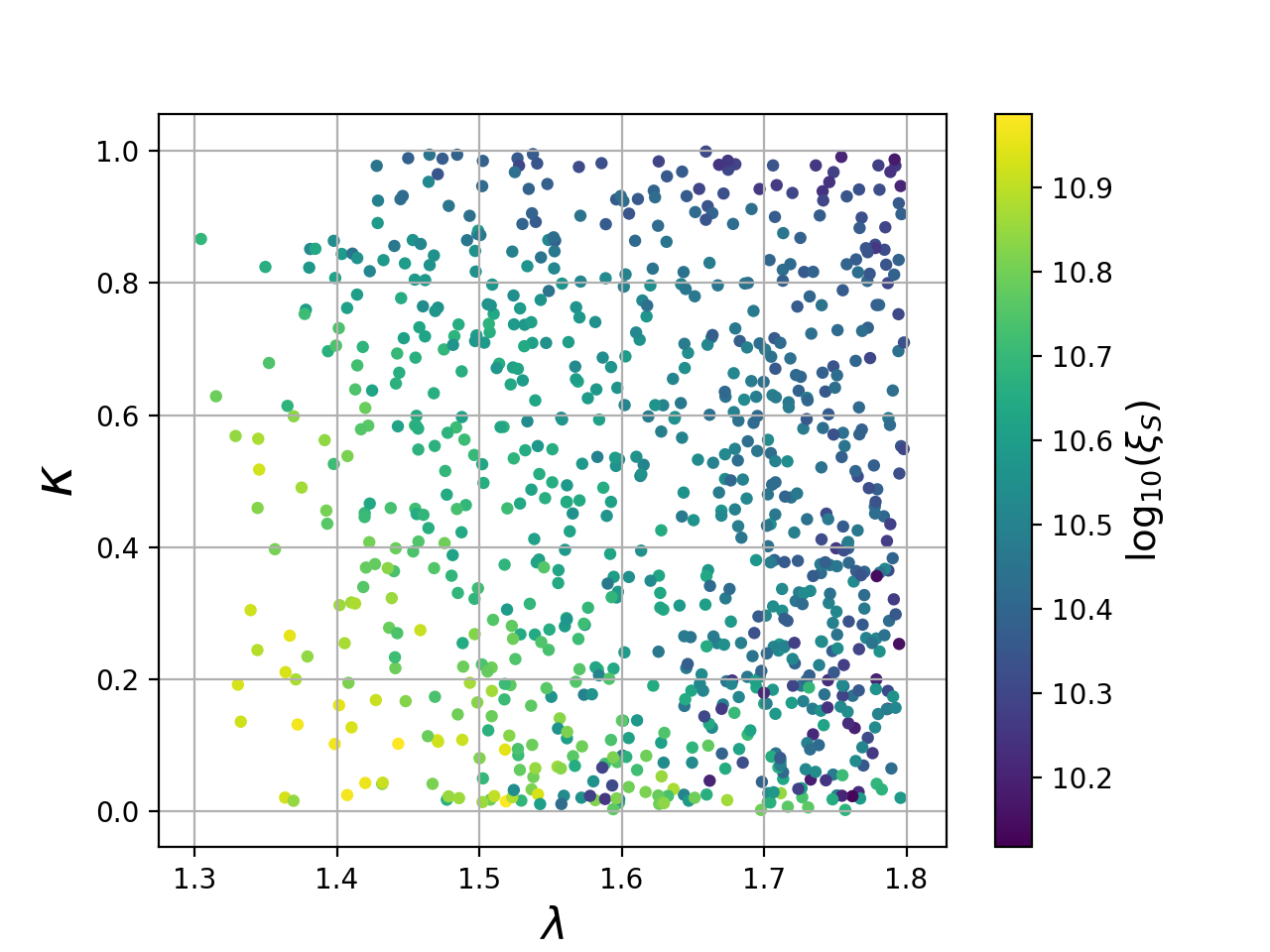}
    \caption{Scatter plot of points that survive the $125$~GeV mass constraint and predict a wrong-sign bottom Yukawa coupling. The colorbar on the upper plot shows the value of $\kappa_b$, which is the ratio between Higgs to $b\bar{b}$ coupling and its SM value, i.e. $g^{NMSSM}_{hb\bar{b}}/g^{SM}_{hb\bar{b}}$. All points have $\kappa_b$ close to -1 as demanded. The lower plot shows the relationship between the values of $\lambda$, $\kappa$, and the tadpole contribution.}
    \label{NMSSM}
\end{figure}

In the rest of this section, we examine the results of a  scan to search for models which produce a wrong-sign bottom Yukawa coupling. Based on the above analysis, we fix 
the supersymmetry mass parameters to the values given above and vary the tadpole term $\xi_S$ in the range of $1.25\times10^{10}$ and $1\times10^{11}$ GeV$^{3}$ and $\mu$ in the range $-1300$ to $-800$ GeV. Beyond these ranges, the number of successful points decreases quickly due to experimental limits. For larger $\xi_{S}$ and $\left| \mu \right|$, the charged Higgs mass becomes too low; for lower $\xi
_{S}$ and $\left| \mu \right|$, the range of $\xi_{S}$ and $\mu$ which passes experimental constraints and gives the correct SM-like Higgs mass becomes quite narrow, and $\kappa_b$ can become positive for $\mu$ close enough to 0. We therefore do not include these ranges in the main scan. Additionally, the following parameters were varied within the ranges of $t_\beta \in [6, 10]$, $M_{A} \in [300, 400]$ GeV, $\lambda \in [1.0, 1.8]$, $\kappa \in [0.0, 1.0]$.
Parameters are randomly drawn from uniform distributions and we discard all points which give the wrong lightest Higgs mass or fail other collider direct experimental constraints
as defined in NMSSMTools. We did not impose any flavor or Dark Matter constraints, but we shall discuss these constraints in separate sections. The results of this analysis are shown in Fig.~\ref{NMSSM}. 

It is clear that the requirement of a wrong-sign bottom Yukawa indeed fixes $\lambda$ to be of order 1, a reflection of the strong tree-level dependence of the bottom coupling on this parameter. Additionally, larger values of $\xi_{S}$ allow $\lambda$ to take on lower values; however, as shown in Fig. \ref{NMSSM_tadpole}, $\xi_{S}$ is also restricted from the requirement of obtaining the proper Higgs mass and cannot take arbitrary large values. Therefore, it is difficult to push $\lambda$ down below order 1 using the tadpole contribution. The constraint on $\kappa$ is significantly relaxed by the large value of $\xi_{S}$, which allows $\kappa$ to take on values from $0$ to $1$, with a gentle dependence on $\lambda$ for given $\xi_{S}$. Further parameters and relevant outputs for five sample points which were successful are provided in Table 1. It is clear from Table 1 that in this region of parameters $A_\lambda \ll m_S$ and therefore the effects associated
with the singlet decoupling, Eq.~(\ref{eq:singdec}), become small.  As we shall see in later sections, this situation will change when we consider values of $\xi_F \neq 0$. 

\begin{table}[H]
    \centering
    \caption{Typical parameters found by NMSSMTools that gave negative Higgs to $b\bar{b}$ couplings}
    \begin{tabular}{c|ccccccc|cccccc}
    \hline\hline
    $No.$ & $t_\beta$ & $M_A \:$ & $\: \mu \:$ & $\: \: \: \lambda \: \: \:$ & $\: \: \: \kappa \: \: \:$ & $\: A_{\lambda} \:$ & $\: \xi_{S}$ & $\kappa_b$ & $BR(h\to b\bar{b})$ & $m_{h} \:$ &  $\: m_{H} \: $  &  $m_{H^{\pm}}$  & $m_{S}$\\
    \hline
    1  & 9.7 & 374 & -1283  & 1.41 & 0.024 & 11.0 & $9.79\times10^{10}$ & -0.98 & 64.9\% &123.1 & 278 & 159 & $10360$ \\
    2  & 8.5 & 398 & -1294  & 1.37 & 0.131 & 109.3 & $9.38\times10^{10}$ & -0.90 & 63.2\% & 122.3 & 271 & 158 & $9973$ \\
    3 & 7.7 & 369 & -1190  & 1.62 & 0.063 & 31.5 & $6.27\times10^{10}$ & -0.98 & 58.6\% & 127.1 & 310 & 158 & $9242$ \\
    4 & 8.5 & 362 & -1119  & 1.41 & 0.398 & 302.3 & $6.92\times10^{10}$ & -0.97 & 58.9\% & 126.5 & 277 & 156 & $9344$\\
    5 & 8.9 & 331 & -1109  & 1.37 & 0.200 & 150.4 & $7.51\times10^{10}$ & -0.89 & 56.3\% & 125.9 & 273 & 159 & $9634$ \\
    \hline
    \end{tabular}
    \label{table:NM} % is used to refer this table in the text
\end{table}

\begin{figure}[H]
    \centering
    \includegraphics[width=.45\textwidth]{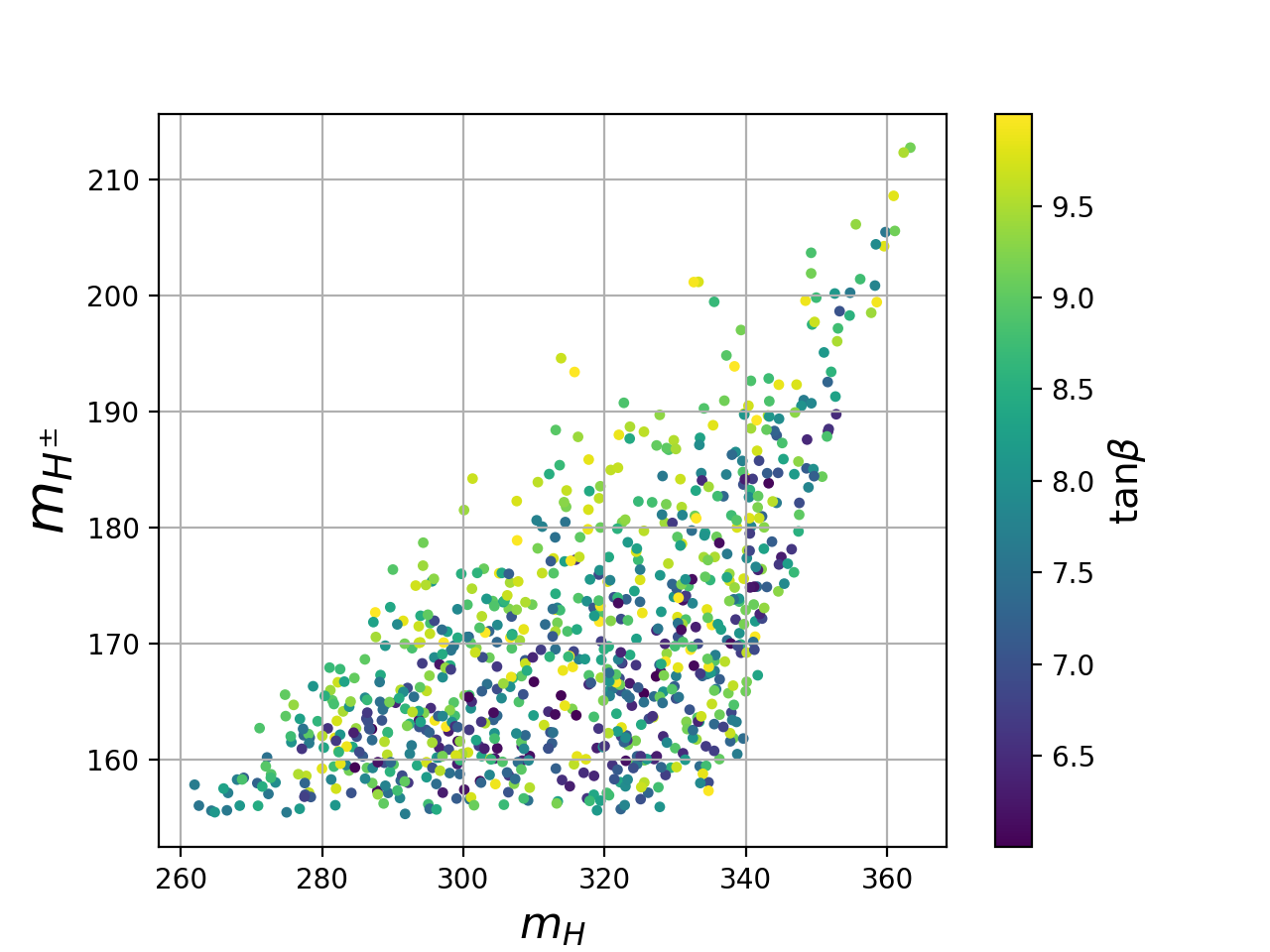}
    \includegraphics[width=0.45\textwidth]{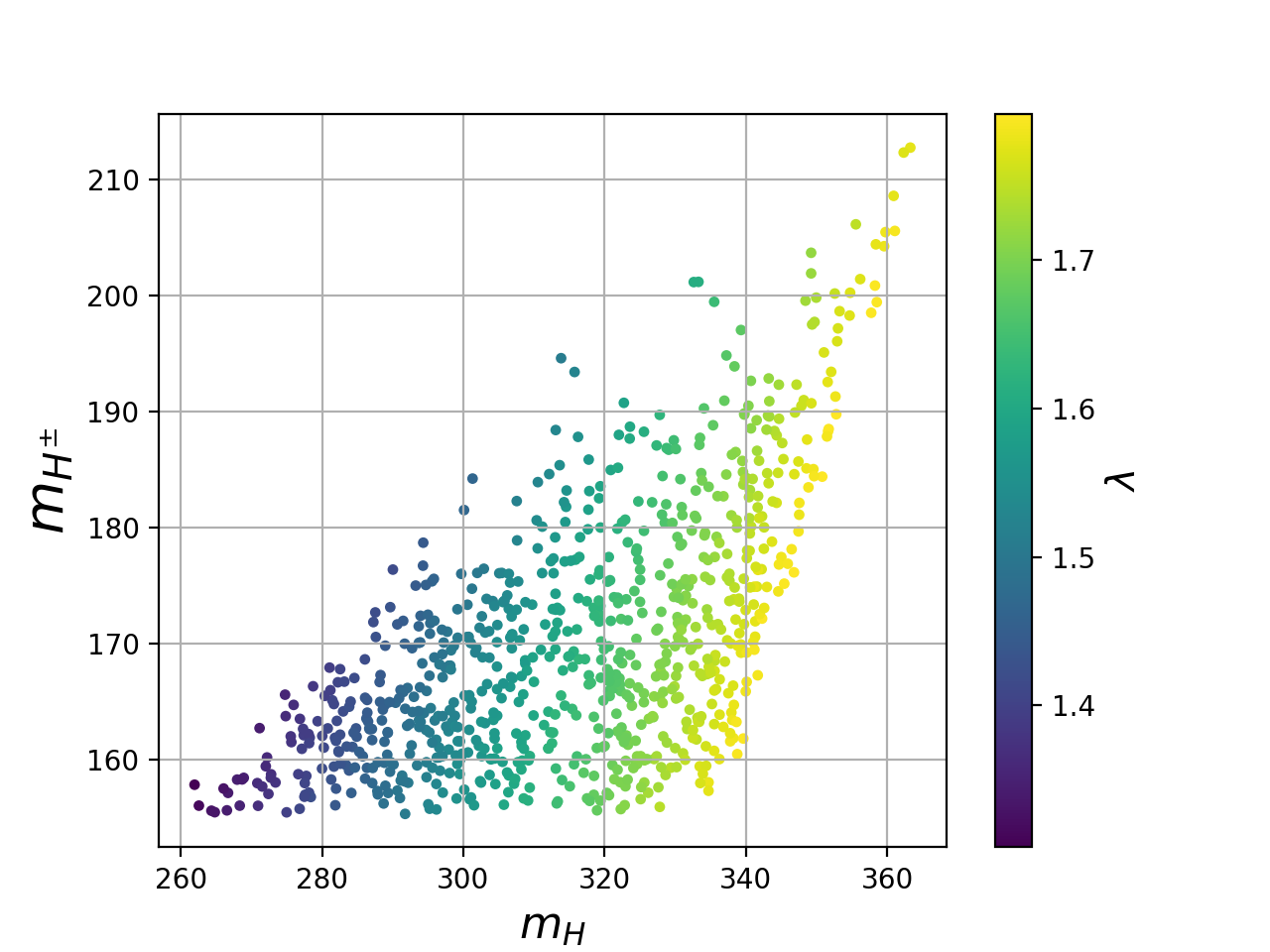}
    \caption{Scatter plot of the charged Higgs mass $m_{H^{\pm}}$ against $m_{H}$, with the colorbar showing the value of $t_\beta$ (left) and $\lambda$ (right). A $t_\beta$-dependent mass cut on $m_{H^{\pm}}$, with a lowest limit of 155 GeV, has been applied to satisfy experimental constraints.}
    \label{MH+}
\end{figure}

All points shown above pass the experimental limits included in NMSSMTools. Additionally, an approximately linear, $t_\beta$-dependent cut is applied to $m_{H^{\pm}}$ based on the constraints provided by CMS \cite{CMS:2014cdp}; a plot of the charged Higgs mass as a function of $m_{H}$ is shown in Fig.~\ref{MH+}. These mass ranges allow for enhanced $H \to H^{\pm} W^{\mp}$ and $A_{1} \to H^{\pm} W^{\mp}$ decays, which will be discussed further in Sections \ref{H decays} and \ref{Heavy Higgs}.

\section{Implications for LHC Physics} \label{LHC}

\subsection{SM-like Higgs boson properties}

The change of sign of the bottom coupling has implications for the loop-induced coupling of the SM-like Higgs to gluons and photons
and is also correlated with changes to the couplings of non-standard Higgs doublets to third generation quarks. 

%There is a bump in this distribution along the straight line $\kappa/\lambda=0.5$, which is related to the Log terms $L_{\mu,\nu} = \log(\frac{\max(4 \nu'^2, \mu^2)}{M_{SUSY}^2})$ and $L_{max1} = \log(\max(\frac{\max(\mu^2, m_Z^2)}{\max(4 \nu'^2, m_Z^2)}, 1))$,  that appear in the chargino and neutralino loop corrections,  with $\nu' = \kappa v_s$ and $\mu = \lambda v_s$.    This max function has clearly a division over the line $\kappa/\lambda=0.5$. \\

\begin{figure}[!htp]
    \centering
    \includegraphics[width=1.\textwidth]{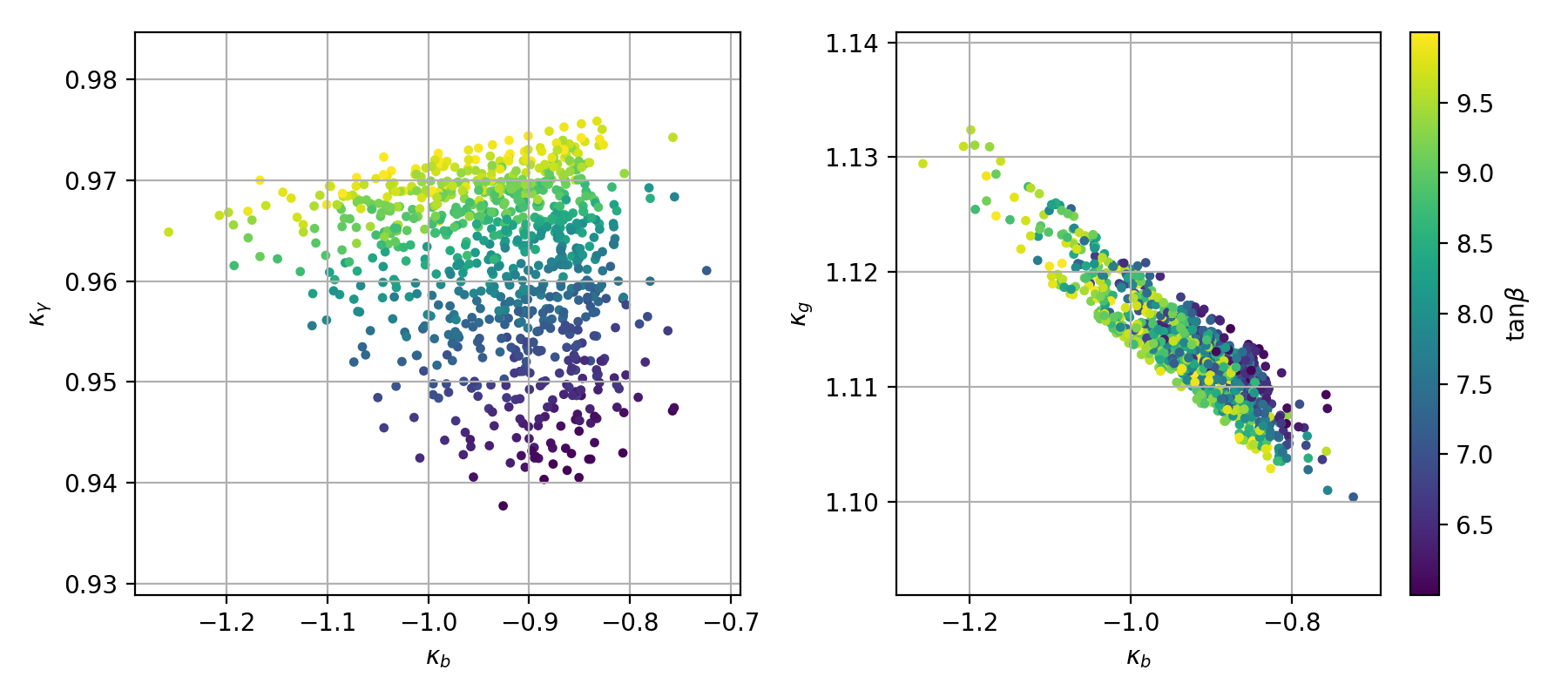}
    \caption{Scatter plots of the couplings for the SM-like Higgs to $\gamma$ (left) and gluons (right) against $\kappa_{b}$. The color bar indicates the value of $t_\beta$. We find that $\kappa_{\gamma}$ is reduced by 3-6\% and displays a linear dependence on $t_\beta$, while $\kappa_{g}$ is enhanced by approximately 10-13\%.}
    \label{NMSSM_couplings}
\end{figure}

In Fig. \ref{NMSSM_couplings}, we plot the values of  $\kappa_g~$ and $\kappa_\gamma~$ against $\kappa_b$, where $\kappa_i$ is the ratio of the Higgs coupling to the particle $i$ to its value in the SM. The $h \to gg$ and $h \rightarrow \gamma \gamma$ amplitudes have contributions from bottom quark loops, and will therefore be modified within our models.  All solutions show values of the couplings that are within 20\% of the SM values, which are in agreement with current experimental constraints. These results
coincide with those obtained  by the authors of Ref.~\cite{Ferreira:2014naa}.

 As can be seen in Fig. \ref{NMSSM_couplings}, the values of $\kappa_{g}$ for our set of points range between 1.10 and 1.13. Since we have assumed heavy supersymmetric particles, these modifications are governed by just the modifications of the bottom couplings.
 This is a reasonably large effect, but observing this effect at the LHC is complicated by systematic errors in the primary $gg$ fusion production cross section. Ref. \cite{Dawson:2013bba} provides expected error estimates for $\kappa_{g}$ of 6-8\% for an integrated luminosity of 300 fb$^{-1}$ and 3-5\% for an integrated luminosity of 3000 fb$^{-1}$. %Any effects from the bottom Yukawa coupling are therefore unlikely to be resolvable from systematic errors 
It is clear from these numbers that hints may become observable by the end of Run 2 and the effects should be clearly resolvable by the end of Run 3.  

The value of $\kappa_{\gamma}$ within our set of points ranges from approximately 0.94 to 0.98. Estimates for LHC uncertainties in the measurement of $\kappa_{\gamma}$ are given as 5-7\% for 300 fb$^{-1}$ integrated luminosity and 2-5\% for 3000 fb$^{-1}$ integrated luminosity \cite{Dawson:2013bba}. The measurement of $\kappa_{\gamma}$ may therefore allow an examination of the viability of the wrong-sign bottom Yukawa within the NMSSM by the end of LHC Run 3.  

Let us stress again that the above estimates of the modification of the Higgs couplings to gluons and photons have been performed under the assumption of heavy supersymmetric particles. If, eventually, charged and/or colored supersymmetric particles are detected at the LHC, their effects would have to be taken into account (see, for instance, Refs.~\cite{Djouadi:1998az},\cite{Dermisek:2007fi},\cite{Carena:2011aa},\cite{Carena:2013iba}) in order to determine the possible effects of the inversion of
the bottom coupling.  

The modification of the SM-like Higgs coupling to top-quarks and weak gauge bosons tend to be small  in the explored region of parameters.  Indeed, ignoring for simplicity the $\Delta_b$ corrections,
\begin{eqnarray}
\kappa_W & = & s_{\beta-\alpha} \simeq 1 -  \frac{2}{t_\beta^2}
\nonumber\\
\kappa_t & = & s_{\beta-\alpha} + \frac{c_{\beta-\alpha}}{t_\beta}  \simeq 1,
\end{eqnarray}
where we have used the fact that $c_{\beta-\alpha} \simeq 2/t_\beta$.

\begin{figure}[!htp]
    \centering
    \includegraphics[width=.6\textwidth]{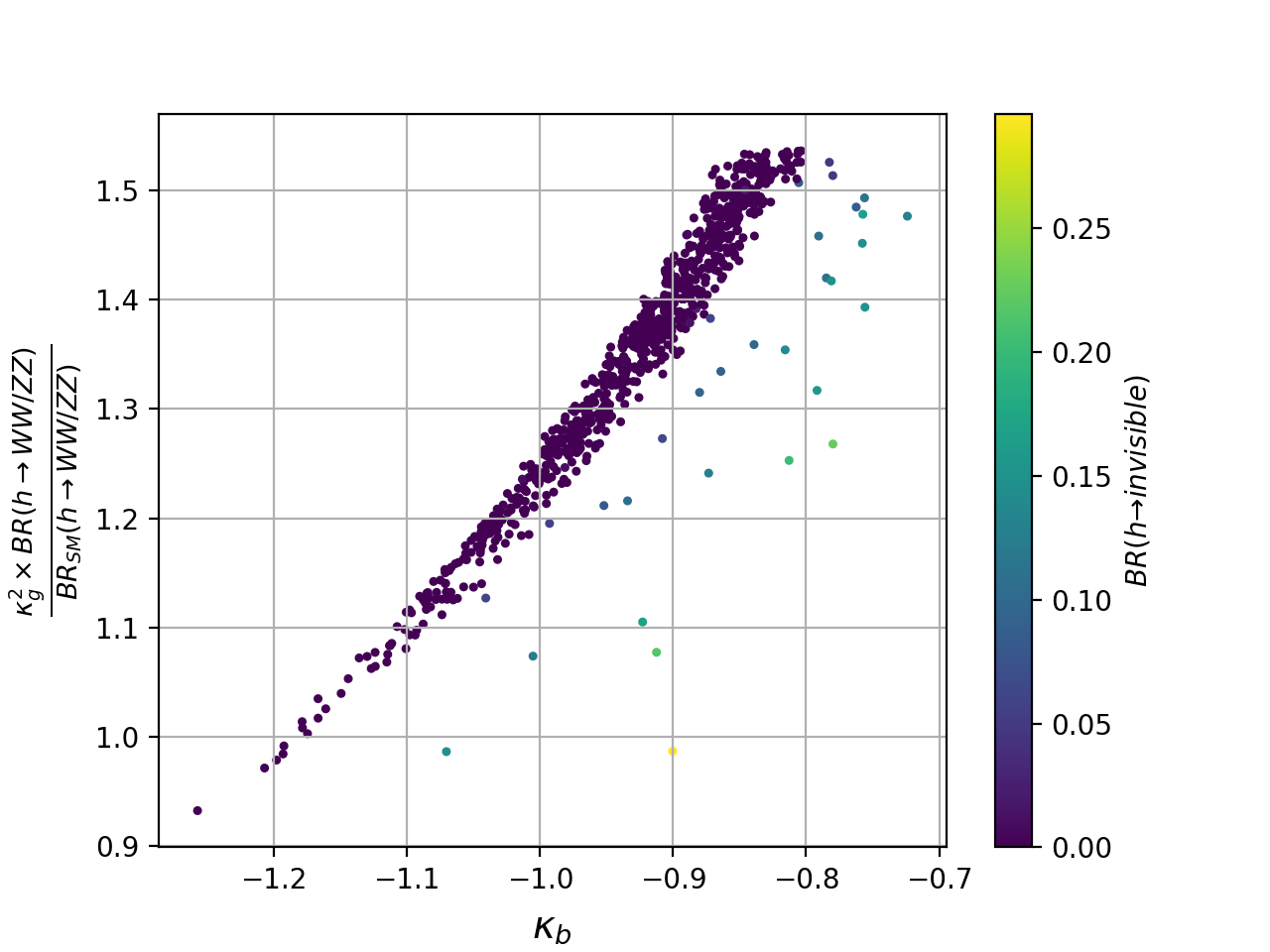}
    \caption{Plot showing the product of $\kappa_{g}^{2}$ and the branching ratio of $h$ to $WW$ or $ZZ$ against $\kappa_b$. The $h \to WW/ZZ$ rates are normalized to the SM rate for the particular SM-like Higgs mass. The colorbar shows the branching ratio of the SM Higgs to neutralinos; we see that the points which do not follow the linear trend have a larger branching ratio to invisible particles.}
    \label{WW_ZZ}
\end{figure}

In Fig.~\ref{WW_ZZ} we show the correlation between the Higgs-induced weak diboson production cross section and the coupling of the SM-like Higgs decay into bottom quarks,
normalized to the values obtained for a Higgs of the same mass in the SM. 
The strong correlation may be explained by the fact that the $BR(h \to WW,ZZ)$ is mostly determined by the variation of the total width induced by the modification of the bottom-quark coupling to the Higgs and by the values of $\kappa_g^2 \simeq 1.25$ (see Fig.~\ref{NMSSM_couplings}). The outlier points which do not follow this linear relationship are
associated with small values of $\kappa$, for which the SM-like Higgs boson can decay into the lightest neutralino and have therefore a non-vanishing branching ratio of decays into invisible particles. 
% If these results are compared with the corresponding values of the signal strength in these channels obtained by the run I combination of CMS and ATLAS 
%data~\cite{Khachatryan:2016vau}
%\begin{equation}
%\mu_F^{ZZ} = 1.42^{+0.37}_{-0.33}  \;\;\;\;\;\;\;\;\;\;\;\; \mu_F^{WW} = 0.98^{+0.22}_{-0.20},
%\end{equation}
%one observes that, although the overall values are acceptable at the 2~$\sigma$ level

\subsection{Radiative Higgs Decay to Quarkonia}

Another particular Higgs process affected by the bottom Yukawa coupling is the radiative decay of the Higgs to Quarkonium, in particular to the $\Upsilon$ meson, which is composed of $b\bar{b}$. This process has also been examined within a general 2HDM in the wrong-sign regime by \cite{Modak:2016cdm}. Within the Standard Model, the direct and indirect Feynman diagrams have an approximate accidental cancelation, which effectively excludes this decay process at all but very high luminosities. The decay widths of $H\rightarrow\Upsilon(nS)+\gamma$ in terms of $\kappa_{b}$ are given by \cite{Bodwin:2014bpa}
\begin{align}
\Gamma[H \rightarrow \Upsilon(1S) + \gamma] &= |(3.33\pm0.03) - (3.49\pm0.15)\kappa_{b}|^2 \times 10^{-10} \text{ GeV}  \nonumber \\
\Gamma[H \rightarrow \Upsilon(2S) + \gamma] &= |(2.18\pm0.03) - (2.48\pm0.11)\kappa_{b}|^2 \times 10^{-10} \text{ GeV}  \\
\Gamma[H \rightarrow \Upsilon(3S) + \gamma] &= |(1.83\pm0.02) - (2.15\pm0.10)\kappa_{b}|^2  \times 10^{-10} \text{ GeV} \nonumber
\end{align}

\noindent where the first term derives from the indirect diagram and the second term, which is modified by $\kappa_{b}$, derives from the direct diagram. Note that the change in sign from $\kappa_{b}=1$ to $\kappa_{b}=-1$ gives a factor increase of between $10^{2}$ and $10^{4}$ in the decay widths. Using $\Gamma(H)=4.195^{+0.164}_{-0.159} \times 10^{-3}$ GeV \cite{twiki:Higgs}, the Higgs branching ratio to $\Upsilon(1S,2S,3S)+\gamma$ final states for the SM are (0.610, 2.15, 2.44)$\times10^{-9}$. For $\kappa_{b}=-1$, the branching ratios are (1.11, 0.518, 0.378)$\times10^{-6}$,
which are still small but significantly larger than the SM values.

The predicted number of $H \rightarrow \Upsilon(nS) + \gamma$ events at the LHC is calculated as

\begin{equation} N=\frac{ \Gamma(H \rightarrow \Upsilon(nS) + \gamma) }{ \Gamma(H) } \times \sigma(p+p \rightarrow H) \times \mathcal{L}_{int}. \end{equation}

\noindent We calculate the expected number of $H \rightarrow \Upsilon(nS) + \gamma$ events for both $\kappa_{b}=1$ and $\kappa_{b}=-1$. The Higgs total cross section is taken to be $\sigma(p+p \rightarrow H)=5.57 \times 10^{4}$ fb \cite{twiki:Higgs}. We examine the number of expected events by the end of LHC Run 3, for which the approximate target integrated luminosity is 300 fb$^{-1}$ \cite{Bordry:2016}. The predicted number of events are less than 1 for $\kappa_{b}=1$ and $N( \Upsilon(1S), \Upsilon(2S), \Upsilon(3S) ) = (18.5\pm0.7, 8.65\pm0.36, 6.31\pm0.26)$ for $\kappa_{b}=-1$. The number of events at the 3~ab$^{-1}$ high-luminosity LHC is simply an order of magnitude larger than the one predicted at the end of Run 3, namely a few hundred events.

Searches for $h \rightarrow \Upsilon(nS) + \gamma$ have been performed previously for the 8 TeV runs with approximately 20.3 fb$^{-1}$ of luminosity \cite{Aad:2015sda}. The current upper limits on the branching ratios at 95\% CL are given for $\Upsilon(1S, 2S, 3S)+\gamma$ final states as (1.3, 1.9, 1.3)$\times10^{-3}$ (\cite{Contino:2016spe}, \cite{Aad:2015sda}). An increase in sensitivity for these decays on the order of $10^{3}$ with respect to the one at Run~1 is therefore required in order to probe the effects of a wrong-sign bottom Yukawa. Therefore, despite the significant enhancement of the number of events with respect to the SM, this process is not currently an effective method of searching for a wrong-sign bottom Yukawa, and its detection will demand a significant improvement of the current analysis.

\subsection{Decay channels of the heavy neutral Higgs} \label{H decays}

\begin{figure}[!htp]
    \centering
    \includegraphics[width=.65\textwidth]{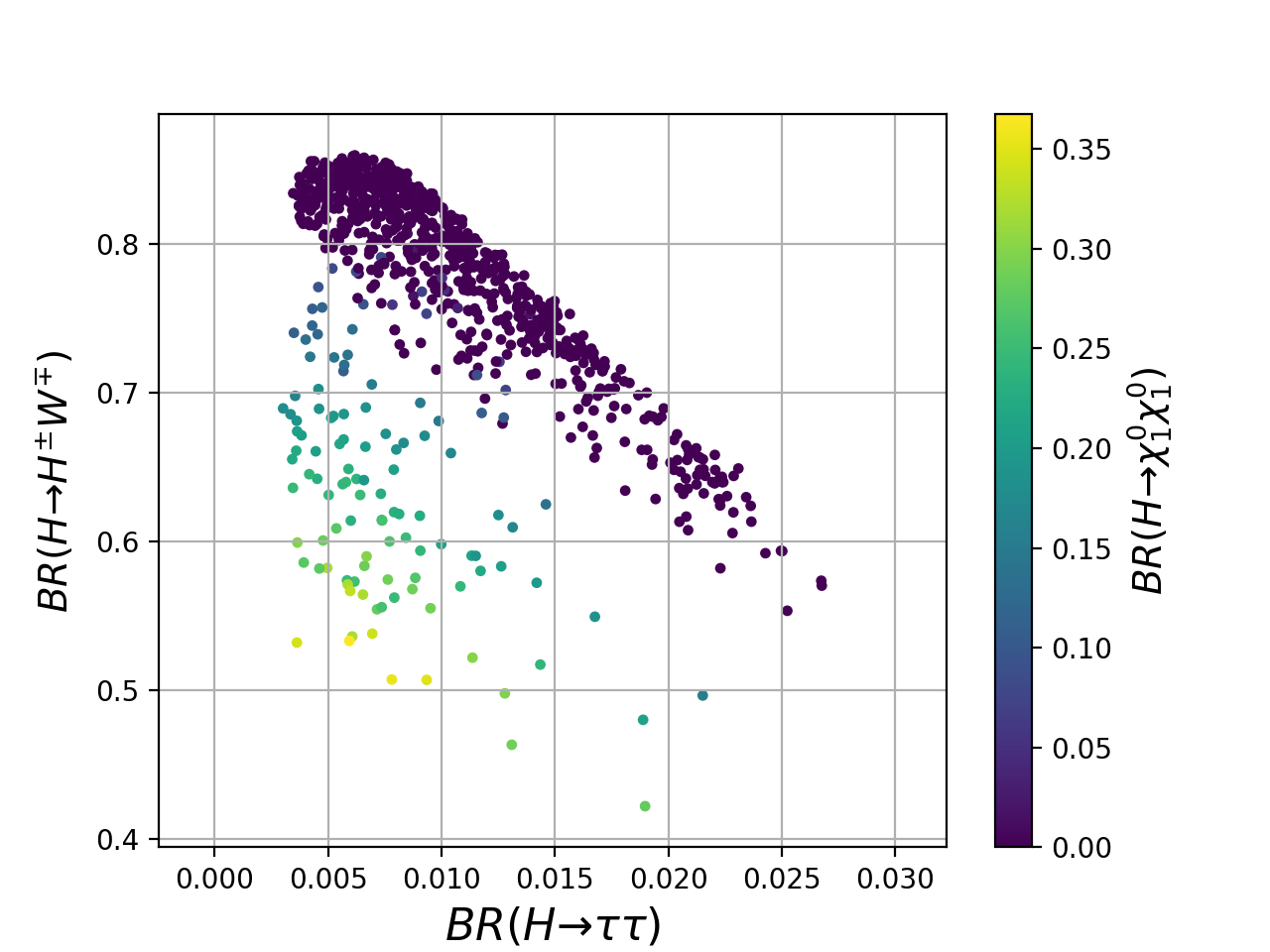}
    \caption{Branching ratios for the decay of the heavier neutral Higgs $H$ to $H^{\pm}W^{\mp}$ and $\tau \tau$, with the branching ratio of $H \to \chi^{0}_{1} \chi^{0}_{1}$ as the colorbar.}
    \label{decay}
\end{figure}

A particular characteristic of those surviving points in Table~\ref{table:NM} is that they all have low charged Higgs mass. The mass difference between $H$, $A_{1}$, and $H^{\pm}$ allows the $H\rightarrow H^{\pm}W^{\mp}$ channel to open up and become the dominant decay mode of the heavier neutral Higgs $H$, as well as of the lighter CP-odd Higgs $A_{1}$. This observation has many phenomenological consequences. On one hand, the branching ratio of $H\rightarrow \tau^{+}\tau^{-}$ is suppressed even when $t_\beta$ is large, so that one may push $t_\beta$ higher than the current bounds on this channel~\cite{Aaboud:2017sjh},\cite{CMS:2017epy}.
These arguments are confirmed by Fig.~\ref{decay}, in which we see that all $BR(H\rightarrow \tau\tau)$ values are lower than $3\%$. On the other hand, this large $BR(H\rightarrow H^{\pm}W^{\mp})$ also means that measurement through this exotic decay channel at the LHC is possible. Within these models, $BR(H \rightarrow H^{\pm}W^{\mp})$ is greater than 0.4 for all models and reaches values up to 0.8.  For low values of $\kappa$, the singlino may become light and, as shown in Fig.~\ref{decay}, the invisible decay branching ratio of the heavy Higgs bosons may become relevant, and imply a further decrease of the decay into $\tau$-leptons.

\subsection{Flavor Constraints}

 As discussed in the previous section, the presence of a light charged Higgs opens new channels for the neutral Higgs decays that
can be searched at the LHC. A light charged Higgs, however, can also induce large corrections to flavor observables, in particular
to the radiative decay of $B$ mesons into strange ones. In type II 2HDM's, the $b \to s \gamma$ rate is indeed highly enhanced 
in the presence of a charged Higgs~\cite{Hewett:1992is},\cite{Misiak:2017bgg}.  In supersymmetric theories, however, this rate 
depends strongly on other contributions coming from supersymmetric particles, and therefore a light charged Higgs cannot be ruled out by these considerations. 
On one hand, there are the contributions coming from the charginos and stops. It is indeed known that in the supersymmetric limit these cancel exactly 
the SM contributions to the dipole operators contributing to the $b \to s \gamma$ transition~\cite{Barbieri:1993av}.  On the other hand, there are flavor 
violating contributions of the neutral Higgs bosons, as well as modifications of the charged Higgs couplings, coming from similar radiative 
corrections to the ones that contribute to $\Delta_b$, discussed in section~\ref{sec:Moderate}~\cite{Ciuchini:1997xe},\cite{Degrassi:2000qf},\cite{Carena:2000uj},\cite{Buras:2002vd}.
All these corrections are included in the NMSSMTools 
code we use~\cite{Ellwanger:2004xm}. Finally, there are contributions that are more difficult to evaluate and come from possible flavor violation in the
scalar fermion sector. Those corrections  are induced whenever there is a misalignment of the basis in which the quark and squark
mass matrix are diagonalized, and lead to large corrections induced by gluino-squark loops~\cite{Gabbiani:1996hi}.  These corrections are induced at the
loop level even if they are not present at tree-level at the supersymmetry breaking scale~\cite{Carena:2008ue}.

In view of the above, we have not considered the flavor constraints in this work.  We have checked, however, that for the solutions we
are presenting the flavor bounds coded in NMSSMTools have a strong dependence on the gluino mass and that small changes to $\mu$ on the order of 10 GeV along with changes of a 
few hundred GeV of the gluino mass, from the 2~TeV value we are considering,  move models from being excluded to being in good agreement 
with flavor constraints. These adjustments leave the behaviors of interest in the Higgs sector unchanged. In addition, as discussed above, the low values of the charged Higgs mass depend strongly on the assumption of
having just a potential tadpole for the singlet. One may push upward the value of the charged Higgs mass with the inclusion of $\xi_{F}$ in the superpotential, which decreases the mass splitting between the charged and CP-odd Higgs. In this case, the dependence on the gluino mass remains and flavor constraints can be satisfied with few hundred GeV adjustments of $M_{3}$.

\section{Heavy charged Higgs} \label{Heavy Higgs}
\subsection{Additional decay channels :  $A_{1}\to hZ$} 

As shown above, models of wrong sign Yukawa couplings have interesting phenomenological properties that go beyond the SM-like Higgs
properties, and include novel decays of the heavy CP-even and CP-odd Higgs bosons that may be tested in the near future. 
ATLAS has recently published results which show an excess of events, consistent with the production of a pseudoscalar resonance of mass
about~400~GeV, produced alongside $b\bar{b}$ and decaying into $hZ$ \cite{ATLAS:2017nxi}. Although one may model this signal with a light singlet \cite{vonBuddenbrock:2017gvy},\cite{vonBuddenbrock:2016rmr}, producing such a pseudoscalar at a high enough rate through $pp \to b\bar{b}A$ production within an effective two Higgs doublet model requires large values of $c_{\beta - \alpha}$ and sizable values of the
bottom-Yukawa coupling, which are consistent with the properties of 
the wrong-sign bottom Yukawa coupling models under study, and is therefore of interest here \cite{Ferreira:2017bnx}. However, one cannot gain an $A_{1} \to hZ$ branching ratio of the necessary magnitude using the minimal models examined above due to the enhanced $A_{1} \to H^{\pm} W^{\mp}$ decay. 

In order to model the $hZ$ decay within these models, we include a non-zero value of the superpotential tadpole term $\xi_{F}$. Because
 $\xi_F$ is a dimension 2 parameter, it is therefore naturally of
the order of $-10^{5}$~GeV$^2$ to $-10^{7}$~GeV$^2$. As noted previously from Eqs. (\ref{eq:MA_xif}) and (\ref{newsplit}), the inclusion of this term reduces the mass difference between the
neutral and charged Higgs bosons and therefore suppresses the decay of the CP-odd Higgs boson into the charged Higgs, increasing
the possible decays into $h$ and $Z$. 

Introducing a non-zero $\xi_{F}$ also allows for larger values of $m_{H}$ and lower values of $\lambda$ while still satisfying $\kappa_{b} \approx -1$.
% The corrections to $\lambda_{7}$ arising from the decoupling of the singlet lead to an increase of $\lambda_7$ for negative values of $A_{\lambda}$.
% In general, from Eqs.~(\ref{eq:condition1}) and ~(\ref{eq:condition2}), it follows that for a given value of $m_H$, these contributions lead to an enhancement of $c_{\beta-\alpha}$  if $\delta\lambda_7 t_\beta + \delta(\lambda_4 + \lambda_5) > 0$. 
As we showed explicitly in Eq. (\ref{eq:tbcba_singlet}), the additional term arising from $\delta\lambda_7$ provides a positive contribution to the value of $t_{\beta} c_{\beta - \alpha}$. 
%while the modifications to $\lambda_4$ and $\lambda_5$ lessen the dependence of the charged Higgs mass on the $\lambda^2$  term, Eq.~(\ref{chhmass}). 
%This would
%hence enable larger values of $m_H$ and lower required $\lambda$ consistent with the condition of negative bottom Yukawa coupling, Eq.~(\ref{eq:wrongsign}).  

\begin{figure}
	\centering
	\includegraphics[width=0.65\textwidth]{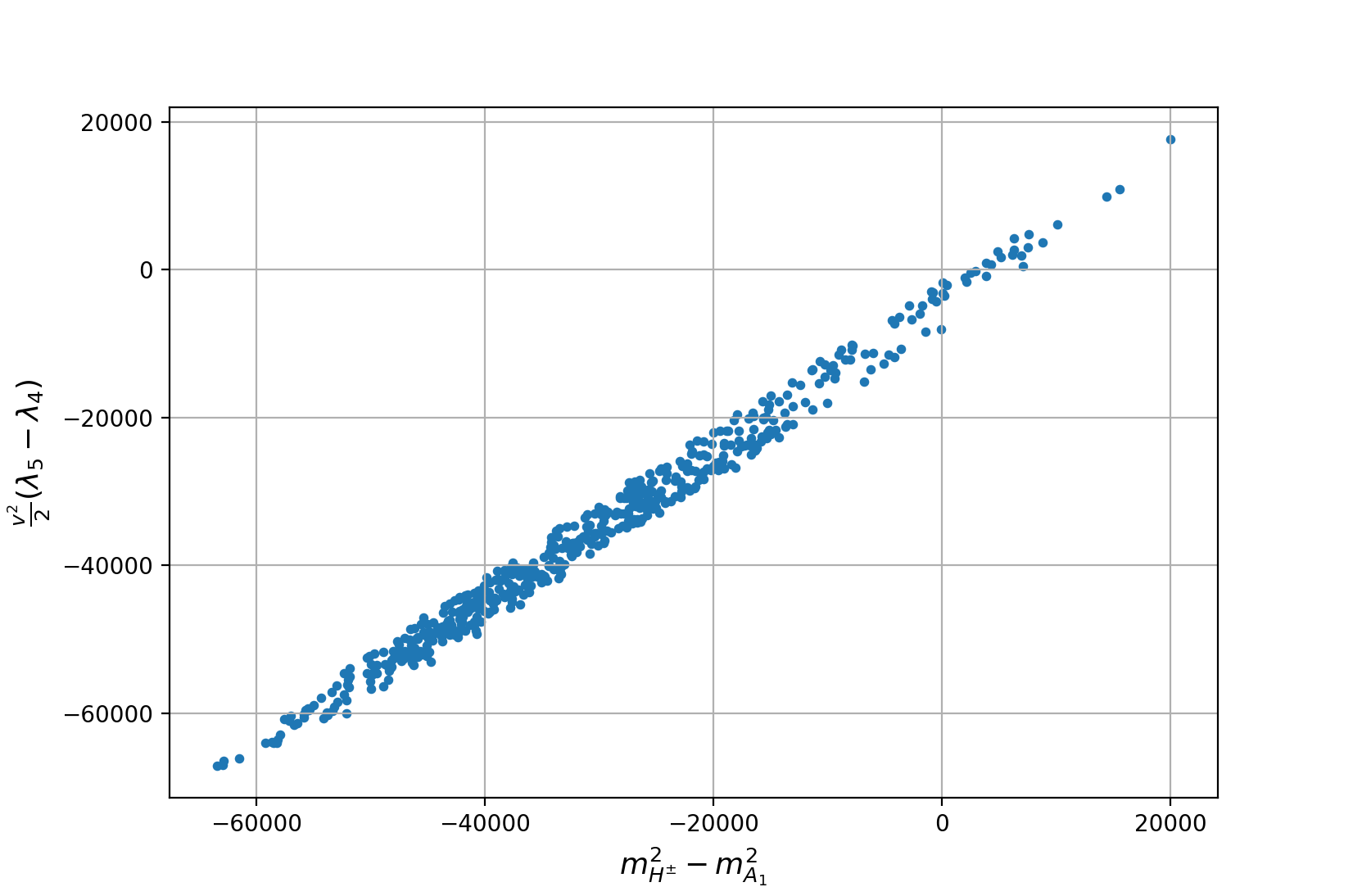}
	\caption{Plot showing the $H^{\pm}$ and $A_{1}$ mass split calculated using our derived expression for $\frac{v^2}{2}(\lambda_5 - \lambda_4)$ against the actual mass split for models with sizeable $\xi_{F}$. There is good agreement between the two values.}
	\label{MassSplitCheck}
\end{figure}

This analysis relies on our approximations of corrections to the $\lambda_{4,5,7}$ couplings; the expressions for $\delta \lambda_{4}$ and $\delta \lambda_{5}$ are verified against the mass splitting $m_{H^{\pm}}^{2} - m_{A}^{2}$ computed by NMSSMTools for large $\xi_{F}$ in Fig. \ref{MassSplitCheck}. We find very good agreement between the actual splitting from our data and the values calculated using the approximations given in Eq.~(\ref{eq:singdec}).

\begin{table}[H]
    \centering
    \caption{Benchmark scenarios for $b\bar{b}$-associated production of $A_1$ decaying into $hZ$. The column ``Rate" represents the quantity $\sigma( pp\to b\bar{b}A_{1} \to hZ) \times BR(h \to b\bar{b})$. All masses are given in GeV.}
    \begin{tabular}{c||ccccccc|cccccccc|}
    \hline\hline
    $No.$ & $t_\beta$ & $\: \: \: \mu \: \: \:$ & $\: \: \: \lambda \: \: \:$ & $\: \: \: \kappa \: \: \:$ & $\xi_{S}$ & $A_{\lambda}$ & $\xi_{F}$ & $\: \: \kappa_b \: \:$ & $\: \: m_{h} \: \:$ & $\: \: m_{H} \: \:$ & $m_{H^{\pm}}$ & $\: m_{S} \: $ & $\: m_{A_1} \:$ & Rate\\
    \hline
    1  & 9.6 & -587  & 1.39 & 0.326 & $3.0\times10^{9}$ & -2779 & $-1.2\times10^6$ & -1.11 & 124.6 & 359 & 384 & $2670$ & 396 & 0.19 \\
    2  & 9.2 & -579  & 1.33 & 0.500 & $2.6\times10^{9}$ & -3157 & $-1.5\times10^6$ & -1.22 & 125.1 & 334 & 411 & $2470$ & 414 & 0.19\\
    3 & 10.5 & -576  & 1.54 & 0.328 & $2.9\times10^{9}$ & -2140 & $-0.8\times10^6$ & -1.15 & 123.0 & 398 & 378 & $2747$ & 421 & 0.18\\
    4 & 8.0 & -784  & 1.45 & 0.405 & $5.9\times10^{9}$ & -3321 & $-1.9\times10^6$ & -1.18 & 123.3 & 351 & 372 & $3325$ & 397 & 0.21\\
    5 & 10.8 & -586  & 1.28 & 0.464 & $3.0\times10^{9}$ & -3345 & $-1.6\times10^6$ & -1.21 & 122.3 & 355 & 426 & $2583$ & 424 & 0.18\\
    \hline
    \end{tabular}
    \label{table:A1} % is used to refer this table in the text
\end{table}

\begin{figure}[H]
	\centering
	\includegraphics[width=0.65\textwidth]{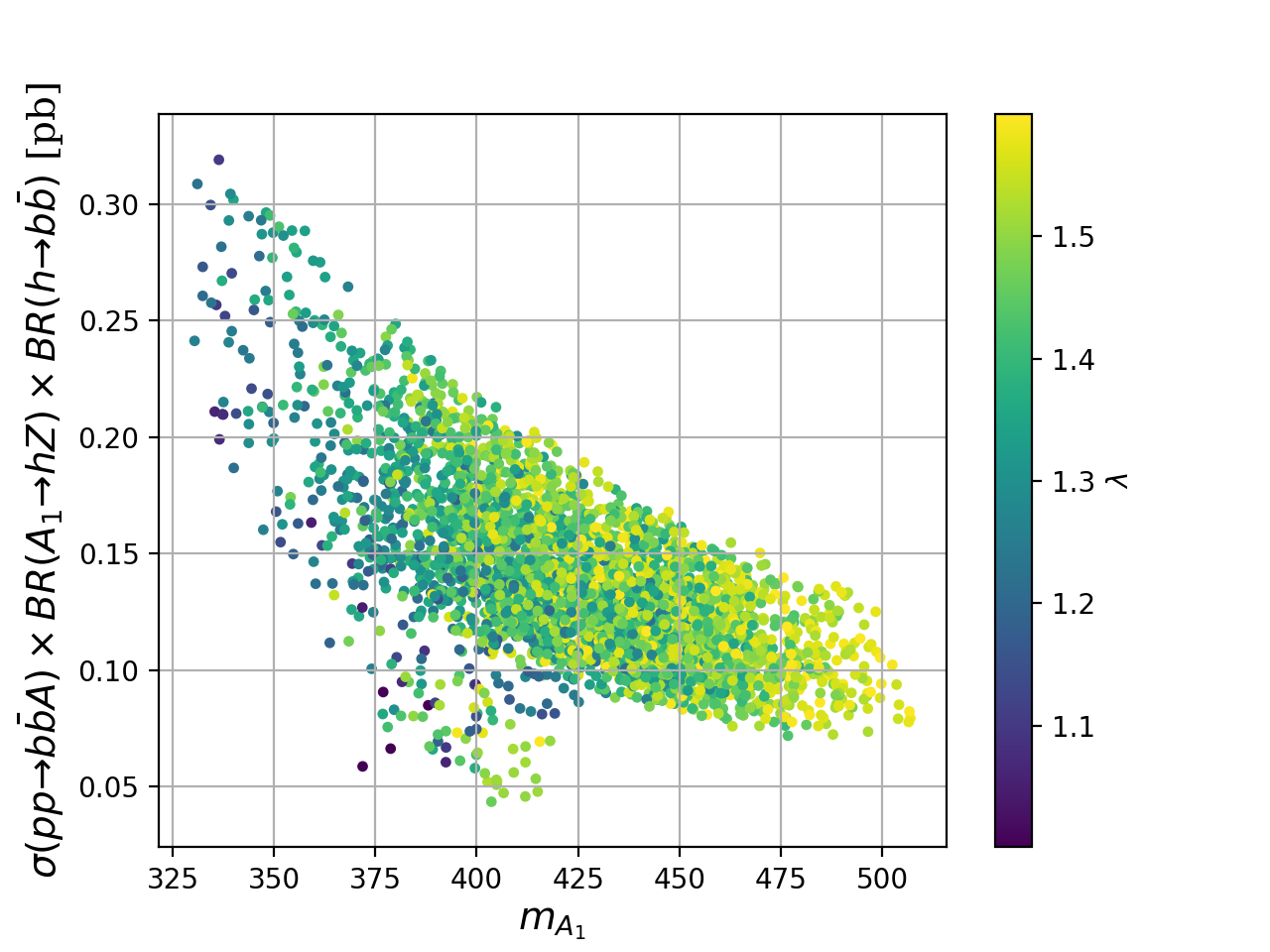}
	\caption{Predicted value of $\sigma( pp\to b\bar{b}A_{1} \to hZ) \times BR(h \to b\bar{b})$ plotted against the mass of the pseudoscalar. The value of lambda for each model is given by the colorbar. We note that one may have a production rate near 0.2~pb for $m_{A_{1}} \approx 400$~GeV. } 
	\label{hZrate}
\end{figure}

%The production cross section $\sigma( pp \to b\bar{b}A_1 )$ is not calculated for each individual model within NMSSMTools, and 
We  calculate the $\sigma( pp \to b\bar{b}A_1 )$ production cross section by scaling the SM cross section by the square of the scaling of the $A_1$ and $b$ coupling relative to the SM value, which is provided by NMSSMTools. The SM cross section scales downward with the Higgs mass, and we fit this dependence by using the SM values provided by the Higgs working group~\cite{deFlorian:2016spz}. 
%User Control Center (FHUCC). 
%We note that while this is not an exact replication of the true dependence, this calculated value for $\sigma$ is within a few percent of the actual predictions from FHUCC, and is therefore a good enough approximation to identify whether our models can produce the current observation. 
The calculated value for $\sigma(pp \to b\bar{b}A_1)$ ranges from 300 to 1600 fb for our particular models, with most falling within the range of 400-800 fb. With these cross section values, we find $\sigma (pp \to b\bar{b}A_{1} ) \times BR( A_{1} \to hZ) \times BR(h \to b\bar{b})$ between 0.05 pb and 0.30 pb. A plot of the predicted rate against the mass of the pseudoscalar is shown in Fig. \ref{hZrate}. We find that these models can approximately produce the observed excess at around 400 GeV, which is currently measured as $\sigma( pp \to b\bar{b}A ) \times BR( A \to hZ) \times BR(h \to b\bar{b}) \approx 0.2$ pb~\cite{ATLAS:2017nxi}. Relevant parameter values which have been changed from the models discussed in the previous section are given by $A_{\lambda} \in [-3500, -2000]$~GeV, $\xi_{S} \in [2.5 \times 10^{9}, 1.6\times 10^{10}]$~GeV$^3$, $\mu \in [-900,-500]$~GeV, $t_\beta \in [8,11]$, $\lambda \in [1.0,1.6]$, $\kappa \in [0.2, 1.0]$, and $M_{A} \in [400,410]$~GeV. Table \ref{table:A1} shows typical parameter values which give a rate for the pseudoscalar production near 0.2 with $m_{A_1}$ near 400 GeV.

\begin{figure}[H]
	\centering
	\includegraphics[width=0.6\textwidth]{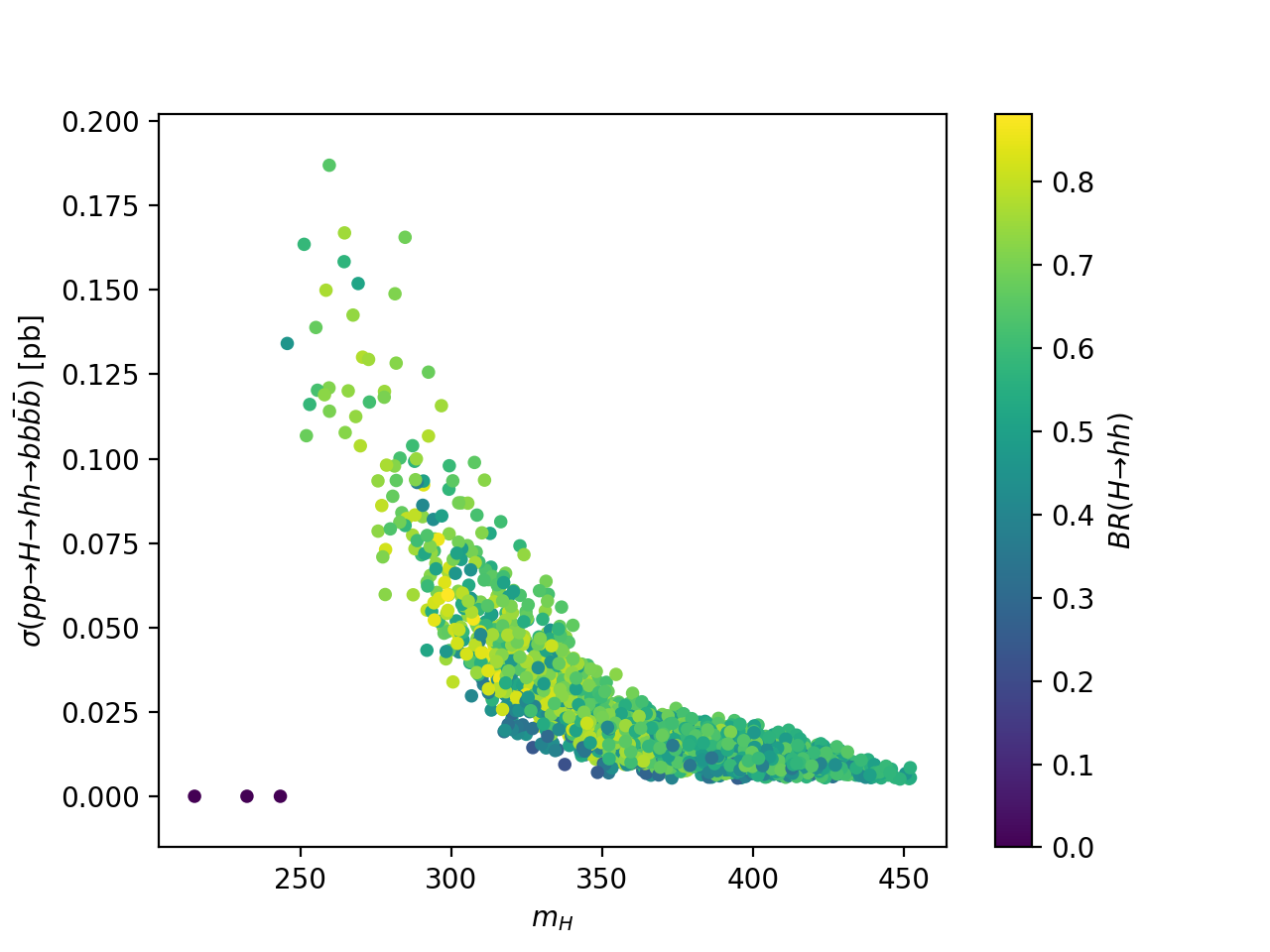}
	\caption{Predicted rate of $H$ production through gluon fusion, decaying into $hh$ and subsequently into $bb\bar{b}\bar{b}$, against the mass of the heavy CP-even Higgs. The colorbar shows the branching ratio of $H \to hh$, which is enhanced in these models. The production rates fall below the current experimental upper limits from CMS.}
	\label{hhrate}
\end{figure}

With an enhanced $A_{1} \to hZ$ decay, one also expects a corresponding enhancement to the related decay $H \to hh$. CMS has updated limits on the production of a spin-0 particle produced via gluon fusion and subsequently decaying into $hh \to bb\bar{b}\bar{b}$~\cite{Aaboud:2017cxo}, and one should check that this enhanced $H \to hh$ process does not exceed these limits. Indeed, the branching ratio $BR(H \to hh)$ ranges mainly between about 0.5 to 0.8 in these models. However, the production rate of $H$ via gluon fusion is suppressed due to relative signs of the $H$ coupling with the top quark and the bottom quark. Within our models, $s_\alpha \simeq c_\beta$ and $c_\alpha \simeq s_\beta \simeq 1$, which differs from the case with $c_{\beta-\alpha}=0$, where instead $s_\alpha \simeq - c_\beta$. In our case, then, the coupling of the heavy Higgs to the top-quark relative to the SM value is given by $\frac{ s_\alpha } { s_\beta } = \frac{1}{ t_\beta}$ as opposed to $\frac{-1}{ t_\beta}$ in the alignment limit. Because the gluon fusion production cross section depends on top and bottom loop contributions, such a change of sign impacts the production rate of $H$ through gluon fusion. In our models, the calculated production rate $\sigma( pp \to H \to hh \to bb\bar{b}\bar{b} )$ falls below the limits given by CMS. Fig. \ref{hhrate} shows the production rate for this process against $m_H$ for each model. Similar conclusions apply to the $H \to ZZ$ channel. 

\subsection{Precision electroweak measurements}
\label{sec:prec}

In the regime of large $c_{\beta - \alpha}$, precision electroweak measurements become a relevant constraint on the parameter space. We therefore calculate the values of the parameters $T$ and $S$ within our models to compare with experimental bounds. Since the singlets are heavy, we can compute the precision measurement observables within the low energy 2HDM effective theory. The expression for $\Delta T$ is given by~\cite{Inami:1992rb},\cite{Chankowski:2000an},\cite{Choudhury:2002qb}
\begin{equation}
\begin{split}
	\Delta T =&  \frac{1}{16 \pi s^2_{W} m_{W}^{2} } \Big( c^{2}_{\beta - \alpha} [ f(m_{A},m_{H^{\pm}}) + f(m_{H^{\pm}},m_{h}) - f(m_{A},m_{h})] \\ 
	&+ s^2_{\beta - \alpha} [ f(m_{A},m_{H^{\pm}}) + f(m_{H^{\pm}},m_{H}) - f(m_{A},m_{H}) ] \Big) \\
	&+ c^2_{\beta - \alpha} \Delta T_{SM}(m_{H}) + s^{2}_{\beta - \alpha} \Delta T_{SM} (m_h) - \Delta T_{SM}(m_h)
\end{split}
\end{equation}
where $s_{W} = \sin(\theta_{W})$ and
\begin{equation}
	f(x,y) = \frac{x^2 + y^2}{2} - \frac{x^2 y^2}{x^2 - y^2} \log \frac{x^2}{y^2} \\
\end{equation}
\begin{equation}
\Delta T_{SM}(m) = \frac{3}{16\pi s^2_{W} m_{W}^{2}} [ f(m,m_{Z}) - f(m,m_{W})] - \frac{1}{8\pi c^{2}_{W}}
\end{equation}
while $\Delta S$ is given by~\cite{Inami:1992rb},\cite{Chankowski:2000an},\cite{Choudhury:2002qb}

\begin{equation}
\begin{split}
	\Delta S =& \frac{1}{12\pi} \Big( c^{2}_{\beta-\alpha} \Big[ \log\frac{m_{H}^{2}}{m_{H^{SM}}^{2}} + \log\frac{m_{h}m_{A}}{m_{H^{\pm}}^{2}} + 2\frac{m_{h}^{2} m_{A}^{2}}{(m_{h}^{2} - m_{A}^{2})^{2}} \\
	&+ \frac{ (m_{h}^{2} + m_{A}^{2}) (m_{h}^{4} + m_{A}^{4} - 4m_{h}^{2}m_{A}^{2}) } {(m_{h}^{2} - m_{A}^{2})^{3}} \log\frac{m_{h}}{m_{A}} \Big] \\
	&+ s^{2}_{\beta-\alpha} [(m_{h} \leftrightarrow m_{H})] - \frac{5}{6} \Big)
\end{split}
\end{equation}

\begin{figure}[h!]
	\centering
	\includegraphics[width=0.45\textwidth]{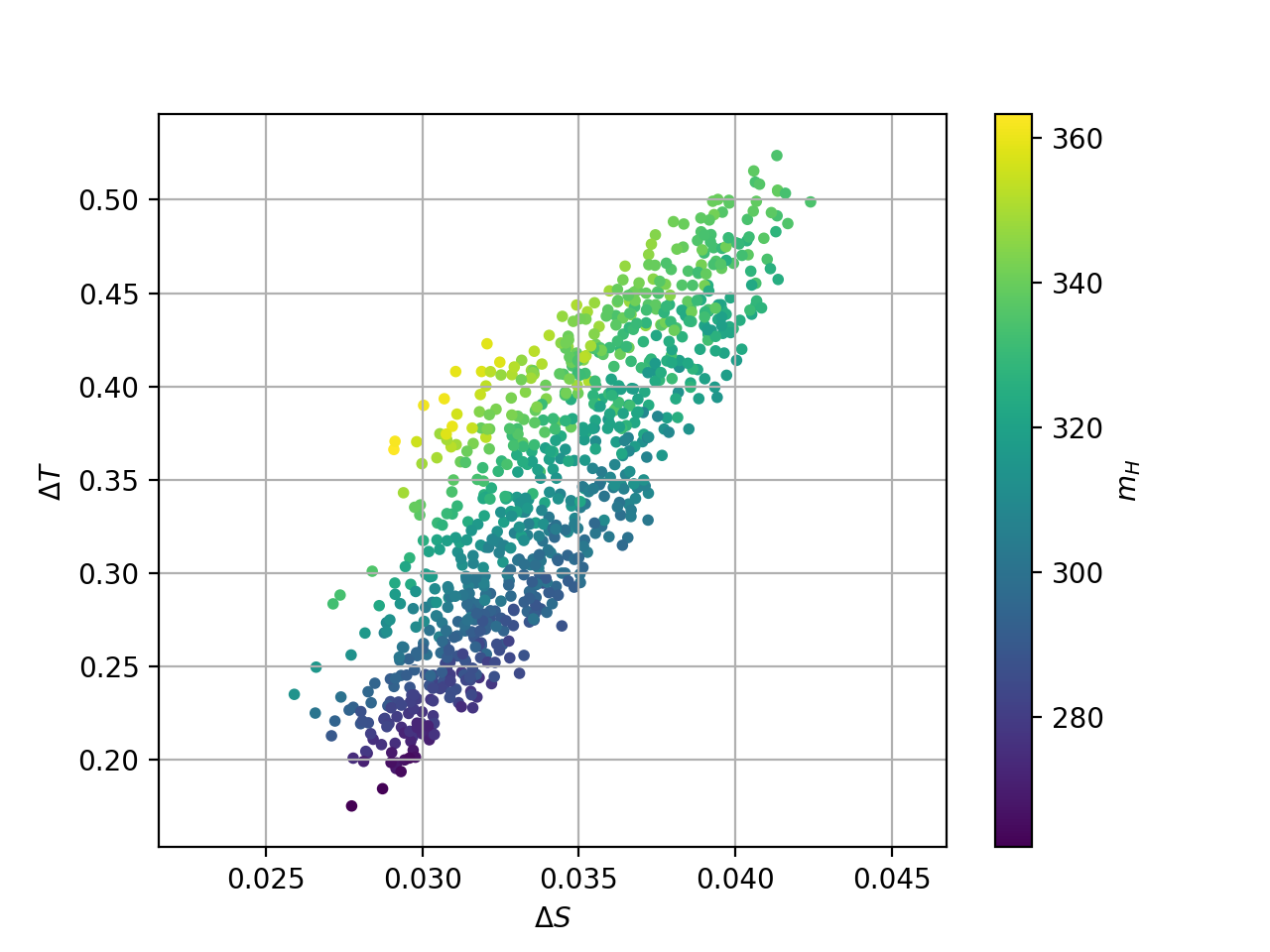}
	\includegraphics[width=0.45\textwidth]{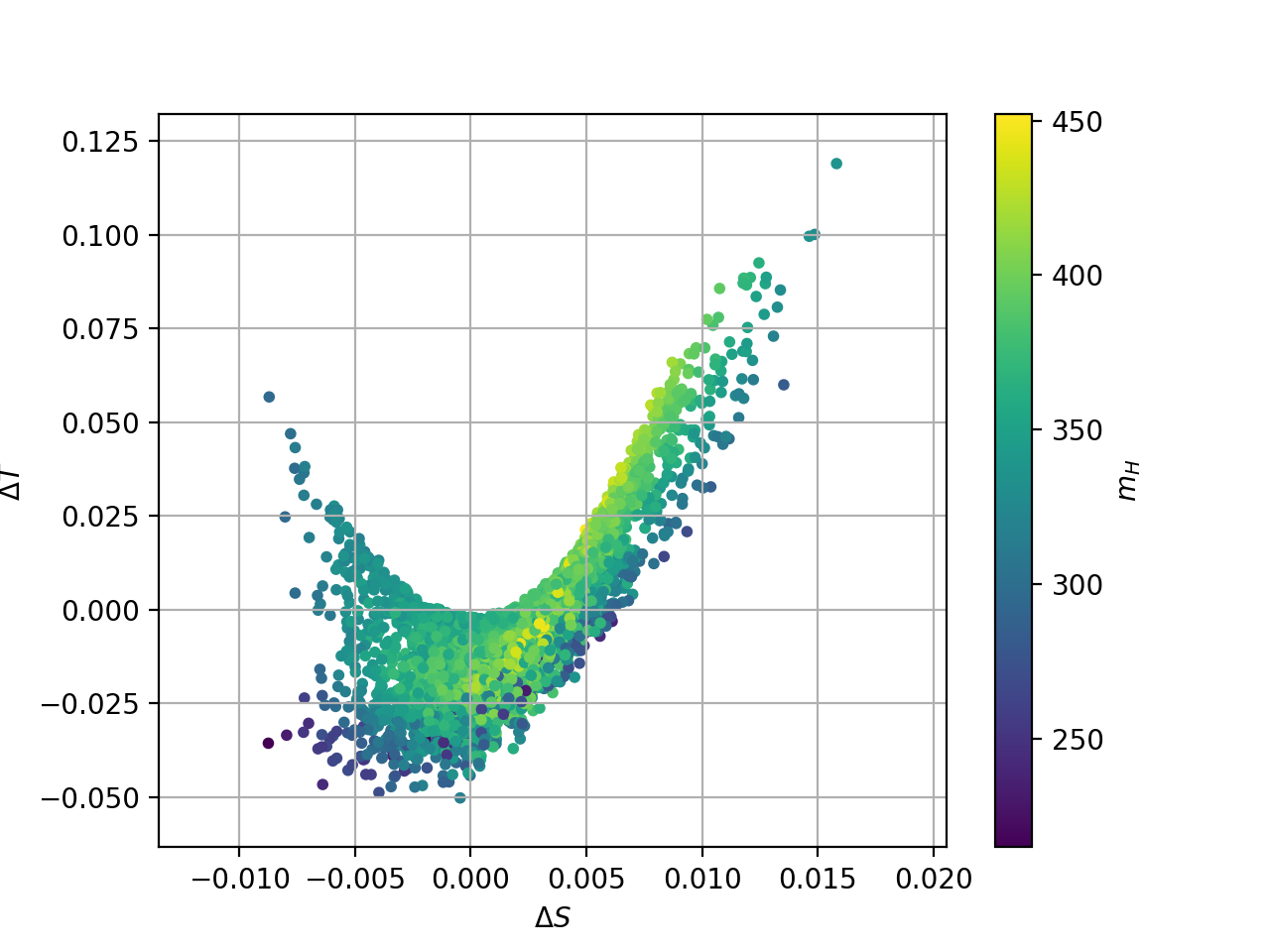}
	\caption{Plots of the precision electroweak parameters $\Delta T$ and $\Delta S$ for the models discussed in Section \ref{Results} (left), with $\xi_{F}=0$, and models with $\xi_{F} \neq 0$ (right). One sees much smaller magnitudes of $\Delta T$ for models with non-zero $\xi_{F}$. For large negative values of $\xi_{F}$, as in the models displayed on the right, the value of $\Delta T$ is well within experimental limits.}
\label{EW}
\end{figure}

Note that due to the custodial symmetry properties, for low splitting between $m_{A}$ and $m_{H^{\pm}}$, the terms $f(m_{H^{\pm}},m_{h,H})$ and $f(m_{A},m_{h,H})$ in $\Delta T$ will approximately cancel; for larger splitting between the masses, i.e. lower $m_{H^{\pm}}$, these terms have a larger contribution. 
%For the $m_{h}$ term, the difference between these two provides a negative contribution; however, the $m_{H}$ term, which is multiplied by $\sin^{2}(\beta-\alpha)$, dominates and provides a positive contribution to $\Delta T$. 
The effects of these variations can be seen in Fig. \ref{EW}. On the left-hand side is a plot of $\Delta T$ versus $\Delta S$ for $\xi_{F}=0$; on the right-hand side is the same plot for models with $\xi_{F} \neq 0$. In the $\xi_{F} \neq 0$ case, the splitting between $m_{H^{\pm}}$ and $m_{A}$ is reduced, as discussed in Section~\ref{NMSSM_singlet}. In this case, we see low values of $\Delta T$. The left-hand plot also shows the dependence of $\Delta T$ on $m_{H}$ in the $\xi_{F}=0$ case, with larger values of $m_{H}$ leading to increased values of $\Delta T$.

In both cases, the value of $\Delta S$ is within the experimental limits. However, for the models presented in Section \ref{Results}, the value of $\Delta T$ exceeds the experimental limits for a number of points. In particular, for the range of $\Delta S \approx 0.035$, the upper limit on $\Delta T$ at 99\% CL is approximately 0.3~\cite{Patrignani:2016xqp}. For $\xi_{F}=0$, one may avoid these constraints by constraining the parameter space to lower $m_{H}$, i.e. $m_H \simlt 320$~GeV; an examination of Fig. \ref{MH+} shows that this corresponds to $\lambda \simlt 1.5$. One may also clearly satisfy these constraints by including a non-zero value of $\xi_{F}$. In light of these results, we conclude that while precision electroweak measurements do provide relevant constraints on the allowed parameter space for the models in Section \ref{Results}, there are a number of existing points which agree with these constraints, and there is additionally a larger class of models which are in good agreement with measurements.

\section{Dark Matter Density and Direct Interaction Cross Section}

The question of Dark Matter in the NMSSM has been investigated by several authors~\cite{Cerdeno:2004xw},\cite{Belanger:2005kh},\cite{Gunion:2005rw},\cite{Cerdeno:2007sn},\cite{Cao:2011re},\cite{Kozaczuk:2013spa},\cite{Han:2017etg}.
In our analysis we have kept the gaugino masses and the Higgsino mass parameter $\mu$ at the TeV scale, implying that, provided $|\kappa| < \lambda/2$, the lightest neutralino is
mostly a singlino with mass
\begin{equation}
m_{\tilde{S}} \simeq 2 \left| \frac{\kappa \mu}{\lambda} \right| .
\end{equation}
As seen in Fig.~\ref{NMSSM}, this condition is fulfilled in most of the parameter space we explored in this article. 
Such a singlino tends to mix with the Higgsino in a relevant way and, due to the large size of the couplngs $\lambda$
and $\kappa$ governing its interactions with the Higgs sector,  the relic density tend to be too small to be consistent
with the experimentally observed one. 

\begin{figure}
	\centering
	\includegraphics[width=0.45\textwidth]{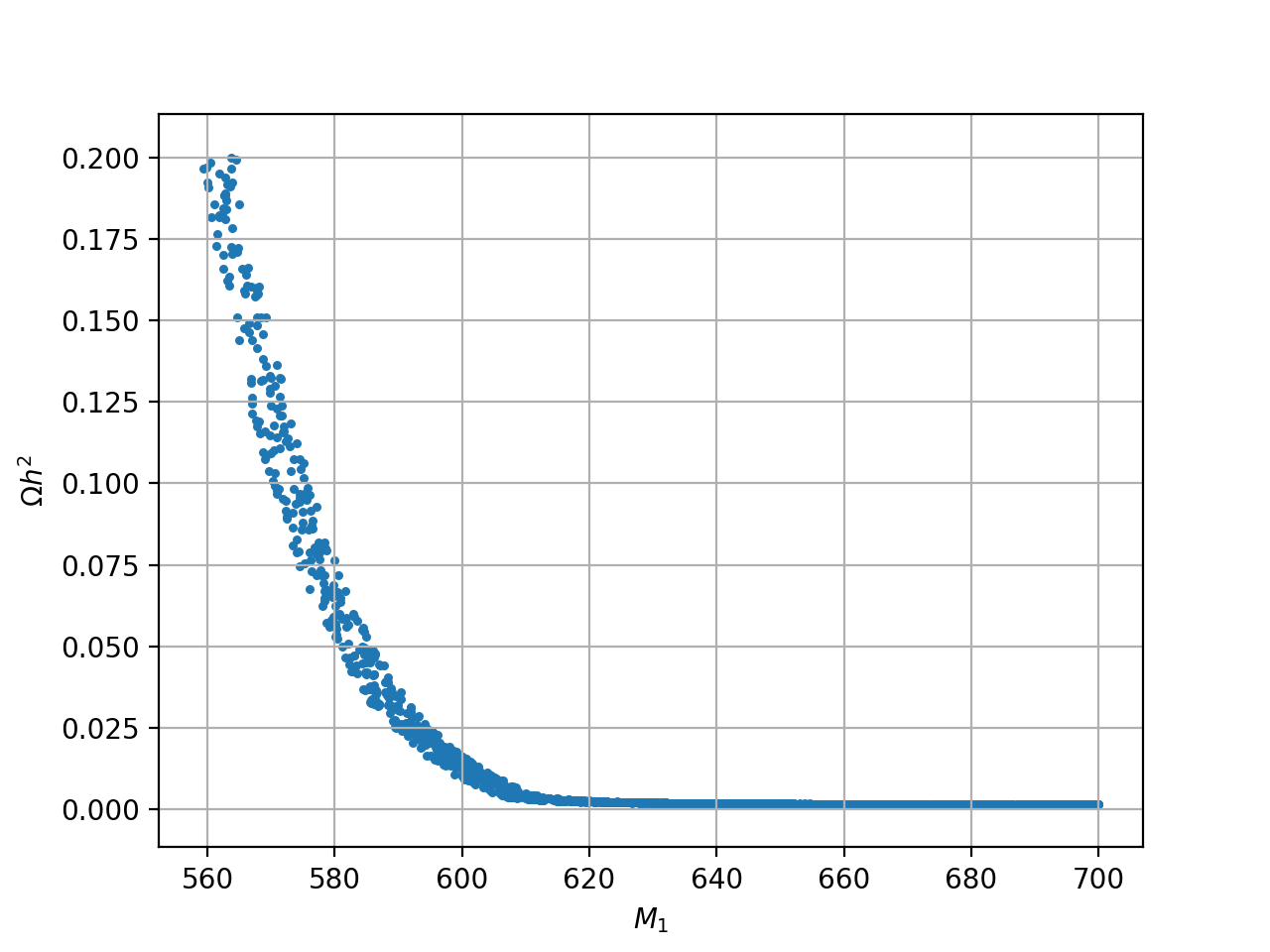}
	\includegraphics[width=0.45\textwidth]{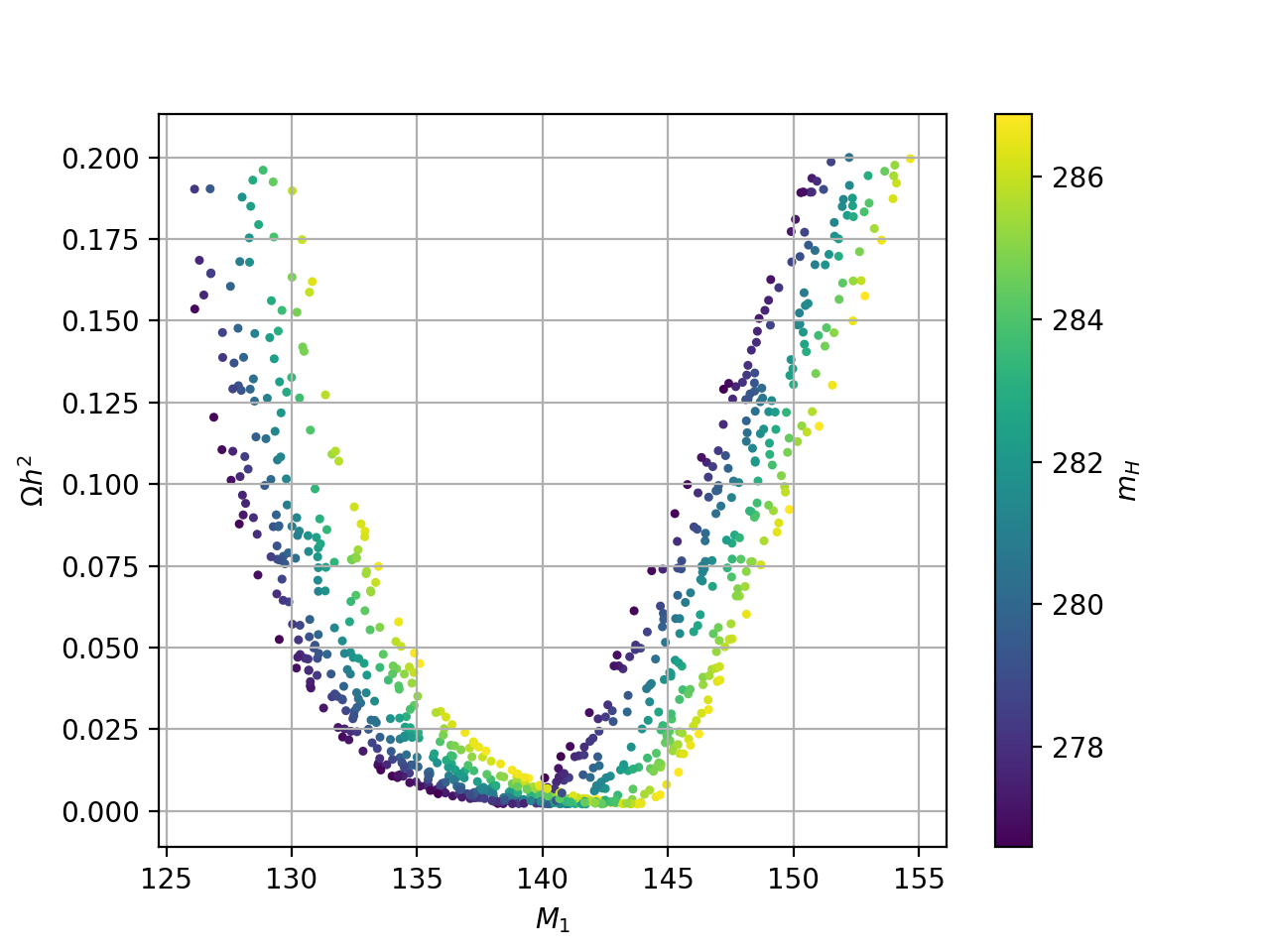}
	\caption{Plots showing the dependence of $\Omega h^2$ on the value of $M_{1}$ for a single model which gives $\kappa_{b}=-1$. The left-hand plot shows the region where $M_{1} \approx m_{\tilde{S}}$, which for this model is about 600 GeV; we see the expected increase in $\Omega h^2$ when $M_{1} \lesssim m_{\tilde{S}}$ due to the lightest neutralino becoming primarily bino-like. The right-hand plot shows the region for which $M_{1} \approx m_{H}/2$, where we see the expected two solutions and strong suppression when $M_{1}$ is about 140 GeV. The widening of the shape of the  plots is due to scanning $\mu$ within a range of 10 GeV, which alters the value of $m_{H}$ by a few GeV. }
	\label{DM_singlepoint}
\end{figure}

Since the relic density could have a different origin from the one associated with the lightest neutralino, a small neutralino component does not
lead to any phenomenological problem. However, it is easy to obtain the observed relic density by modifying the mass parameters and
without affecting the Higgs phenomenology. This may be achieved, for instance,  by lowering the value of the hypercharge gaugino mass
$M_1$.  For low enough values of $M_1$ the lightest neutralino would be Bino-like and the observed relic density could be reproduced under
two circumstances
\begin{itemize}
\item Values of $M_1$ close to a half of the lightest non-standard Higgs masses,  $m_H/2$ or $m_{A_1}/2$~\cite{Griest:1990kh},\cite{Drees:1992am}, 
for which resonant annihilation could take place,
\item Values of $M_1$ close to but lower than $m_{\tilde{S}}$, the so-called well-tempered singlino-bino region~\cite{Baum:2017enm}.
\end{itemize}
If either  of those conditions were fulfilled, not only could the relic density be brought to agreement with the experimentally observed value,
but also the spin-independent and spin-dependent interaction cross section with nuclei will be small enough to be in agreement with
the current experimental constraints.  In our scans, we have modified the values of $M_1$ and verified that this is indeed the case. In particular, for  the values of the parameters present in the benchmark model~4 in Table~\ref{table:NM}, Fig.~\ref{DM_singlepoint} shows the value of $\Omega h^2$ for these two regions of $M_{1}$. For this point, the singlino mass is approximately 600 GeV, while the mass of the heavy CP-even Higgs is about 280~GeV. The widening of the shape of the  plots is due to scanning $\mu$ within a range of 10 GeV, which alters the value of $m_{H}$ by a few GeV. It is clear that as the value of $M_1$ falls below $m_{\tilde{S}}$ and therefore the lightest neutralino becomes primarily bino-like, the relic density increases. On the other hand, for $M_{1}$ near $m_{H}/2$, we see the two regions with $\Omega h^2 \approx 0.1$ on either side of $m_H/2 \simeq 140$ GeV, where the relic density is suppressed by the resonant annihilation.

\section{Conclusions} \label{Conclusions}

The current uncertainties in the determination of the Higgs coupling to bottom quarks leave room for a change of magnitude and sign of
this coupling. In this article we have studied the possible implementation of this idea within the MSSM and the NMSSM. We have
shown that in the MSSM this could only be achieved for values of $m_A$ and $t_\beta$ that are ruled out by current searches for
heavy Higgs bosons decaying into tau pairs. On the other hand, in the NMSSM, consistent solutions that avoid current experimental limits may be found, but for values of the couplings $\lambda$ and $\kappa$ that lead to a Landau pole at scales below the Planck scale.
This perturbativity problem may be solved by either assuming a composite Higgs model or by the introduction of an extended gauge sector that slows down the evolution of $\lambda$ at high energies.

The change of sign of the bottom coupling leads to a modification of the loop-induced couplings of the SM-like Higgs to photons
and gluons that may be tested at higher luminosities at the LHC. In particular, it leads to an enhancement of the order of 20 to 25 percent
of the Higgs production in the gluon fusion mode and a reduction of order of 5 to 10 percent of the width of the decay of Higgs into two
photons. The modification of the gluon fusion production cross section leads already to an enhancement of the Higgs-induced diboson
production cross section, which will allow one to put constraints on the allowed parameter space of the theory. 

The modification of the sign of the bottom coupling also leads to a large enhancement of the radiative decay of the SM-Higgs into
photons and the $\Upsilon$ meson. While such an enhancement leads to a sizable number of radiative decay events at the high
luminosity LHC, the efficiency of the current searches has to be improved in order to lead to an observable signal. 

In this work, we have added tadpole terms to the singlet fields that allow us to raise the value of the scalar singlets and obtain a
realistic scalar spectrum. When only a tadpole for the scalar term is included, 
the required low values of $m_A$ and  large values of $\lambda$ tend to lead to a charged Higgs boson mass that is lower
than the top quark mass, and hence such models are strongly constrained by searches for charged Higgs bosons proceeding from the decay of
top quarks. Models that avoid these constraints have masses of the charged Higgs within 10 to 15 GeV of the top quark mass.
In these models the second lightest CP-even and the lightest CP-odd scalars, which have mainly doublet components, tend to
decay strongly into $H^\pm W^\mp$, which provides an interesting search channel. 

On the other hand, when a tadpole term is also included in the superpotential, the splitting between the CP-odd and the charged Higgs
boson masses may be reduced, suppressing the decay rate of the neutral scalars into charged boson states. In this case, the decay
modes $A_1 \to hZ$ and $H \to hh$ are strongly enhanced. In particular, for values of $m_{A_1}$ of order of 400~GeV, which are naturally
obtained within these models, the production mode $pp \to b \bar{b} A_1 \to b \bar{b} h Z$ may be sizable and can lead to an explanation
of an apparent excess of $hZ$ events at the ATLAS experiment without being in conflict with the current bounds on $H \to hh$ production.

Models with light charged Higgs masses are constrained by flavor and precision measurement constraints. While the flavor constraints
may be avoided by suitable supersymmetric contributions, the precision measurement constraints set a limit on the possible splittings
of the charged and neutral Higgs bosons. Finally, the observed Dark Matter relic density may be obtained by suitable choice of the
gaugino mass parameter $M_1$, without affecting the Higgs phenomenology.

~\\
{\bf ACKNOWLEDGMENT}
We would like to thank the Aspen Center for Physics, which is supported by the National Science Foundation under Grant No. PHYS-1066293, 
and where part of this work has been performed. Work at ANL is supported in part by the U.S. Department of Energy under Contract No. DE-AC02-06CH11357. Work at EFI is supported by the U.S. Department of Energy under Contract No. DE-FG02-13ER41958.

\newpage.

\appendix
\section{Renormalization Group Evolution}

The discussions in previous sections demonstrated that to reverse the sign of the coupling of the Higgs boson to bottom quarks, the $\lambda$ or $\kappa$
couplings need to take sizable values. However, this region of parameters leads to a Landau-pole problem, i.e. coupling constants will reach 
non-perturbative values  at  energies much lower than the GUT scale during the renormalization group evolution (RGE)~\cite{Nevzorov:2001jh}. This problem can be solved
in two ways: the first is assuming that the rise of the quartic coupling $\lambda$ is the evidence of being in the presence of a fat Higgs model~\cite{Harnik:2003rs}, namely
a reflection of the  composite nature of the Higgs fields, which are  just mesons of a confining theory in the UV.  The second is  
by extending the gauge groups, for example, to $SU(3)_c\times SU(2)_1\times SU(2)_2\times U(1)_Y$. More specifically,
one can take  the third generation and Higgs sector to be charged under $SU(2)_1$ while the first two generations
are charged under $SU(2)_2$~\cite{Batra:2004vc}. The symmetry breaking from $SU(2)_1\times SU(2)_2$ to the regular $SU(2)$ is achieved by a bi-doublet chiral field $\Sigma$ at energies
$\left< \Sigma \right> = u$ of the order of a few~TeV.  Large values of the $SU(2)_{1}$ coupling would allow the $\lambda$ coupling to be perturbative up to the
GUT scale. 

While the composite nature of the Higgs fields would be an interesting possibility, which also leads naturally to tadpole contributions to the singlet field $S$, 
we shall present an analysis of to what extent the model can be rendered consistent with perturbation theory up to high energy scales by the addition of extra gauge
interactions.  In order to get a quantitative understanding of the possible modifications of the RGE of the coupling $\lambda$, we performed a one-loop
analysis of the evolution of the couplings to high energies. Taking the new symmetry breaking sector into consideration, above the symmetry breaking scale, and
assuming just the minimal Higgs and gauge particle content to ensure the realization of this model together with approximate unification
of the diagonal $SU(2)_1 \times SU(2)_2$, $SU(3)_c$ and $U(1)_Y$ couplings at the GUT scale, the RGE equations are given by~\cite{Batra:2004vc} 
\bea
\frac{d \tilde{\alpha}_1}{dt}&=& \frac{38}{5} \tilde{\alpha}_1^2,\\
\frac{d \tilde{\alpha}^{(1)}_2}{dt}&=& -2  (\tilde{\alpha}^{(1)}_2)^2,\\
\frac{d \tilde{\alpha}^{(2)}_2}{dt}&=& 4  (\tilde{\alpha}^{(2)}_2)^2,\\
\frac{d \tilde{\alpha}_3}{dt}&=& -2 \tilde{\alpha}_3^2,\\
\frac{d Y_t}{dt}&=&Y_t\left( Y_\lambda + 6 Y_t -\frac{16}{3} \tilde{\alpha}_3 -3\tilde{\alpha}^{(1)}_2- \frac{13}{15}\tilde{\alpha}_1\right),\\
\frac{d Y_\lambda}{dt}&=&Y_\lambda\left( 4 Y_\lambda + 2Y_\kappa+3 Y_t  -3\tilde{\alpha}^{(1)}_2- \frac{3}{5}\tilde{\alpha}_1\right),\\
\frac{d Y_\kappa}{dt}&=&6Y_\kappa\left( Y_\lambda + Y_\kappa\right) ,
\eea
where
\bea
\tilde{\alpha}_i(t) = g_i^2(t)/(4\pi)^2, Y_\lambda(t) = \lambda^2(t)/(4\pi)^2, Y_\kappa(t) = \kappa^2(t)/(4\pi)^2,
\eea
$t = \ln(Q^2)$. In the above, $g_i^{(1,2)}$ correspond to the couplings of the two $SU(2)$ gauge interactions. 

On the other hand, after gauge symmetry breaking, $SU(2)_1 \times SU(2)_2 \to SU(2)_L$ one is naturally left with an effective theory with the same particle content as in
the NMSSM.  Considering only the particles in the NMSSM, we get the following one-loop RGE equations for $\alpha$'s and Yukawa couplings,
\bea
\frac{d \tilde{\alpha}_1}{dt}&=& \frac{33}{5} \tilde{\alpha}_1^2,\\
\frac{d \tilde{\alpha}_2}{dt}&=& \tilde{\alpha}_2^2,\\
\frac{d \tilde{\alpha}_3}{dt}&=& -3 \tilde{\alpha}_3^2,\\
\frac{d Y_t}{dt}&=&Y_t\left( Y_\lambda + 6 Y_t -\frac{16}{3} \tilde{\alpha}_3 -3\tilde{\alpha}_2 - \frac{13}{15}\tilde{\alpha}_1\right),\\
\frac{d Y_\lambda}{dt}&=&Y_\lambda\left( 4 Y_\lambda + 2Y_\kappa+3 Y_t  -3\tilde{\alpha}_2 - \frac{3}{5}\tilde{\alpha}_1\right),\\
\frac{d Y_\kappa}{dt}&=&6Y_\kappa\left( Y_\lambda + Y_\kappa\right)
\eea

\begin{figure}[H]
    \centering
    \includegraphics[width=.7\textwidth]{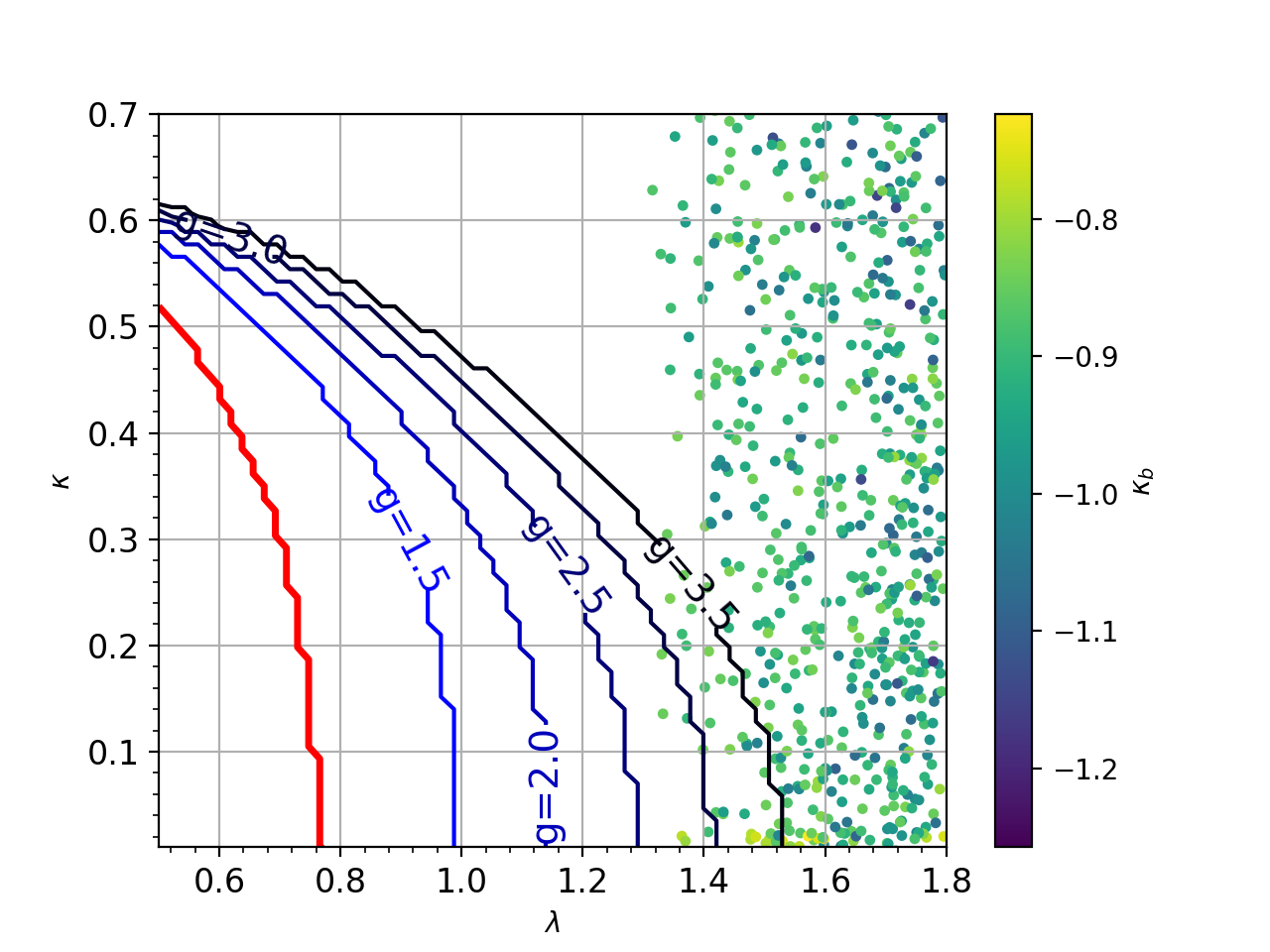}
    \caption{Plot illustrating the Landau pole constraints on $\lambda$ and $\kappa$. Below the red line is the allowed region of $\lambda$ and $\kappa$ in the NMSSM model; the blue contours show the allowed boundaries of $\lambda$ and $\kappa$ for different values of the new $SU(2)$ coupling $g_1$ at $u = 3$~TeV.  Below each line all couplings are perturbative during 1-loop RG evolution up to energies of $10^{16}$ GeV. The contour lines correspond to $g_1 = 1.5, 2.0, 2.5, 3.0,$ and $3.5$ from left to right.  The points show
    the values of $\kappa$ and $\lambda$ associated with negative values of the bottom Yukawa, shown in Fig.~\ref{NMSSM}.}
    \label{new_su2}
\end{figure}

The modification to the renormalization group equations gives us more flexibility in the choice of $\lambda $ and $\kappa$. In Fig.~\ref{new_su2}, we display the RGE result with and without new gauge couplings, plotted along with the successful points for $\kappa_b=-1$ found in Section \ref{Results}. We can see that without new gauge couplings, the constraint from the requirement of avoiding the
Landau-pole problem is quite stringent since the maximum viable value for $\lambda$ is of order 0.7 and becomes smaller for larger
values of $\kappa$.  Therefore, all solutions with negative bottom Yukawa couplings, which are associated with values of $\lambda > 1.0$,  lead to the loss of perturbativity below the GUT scale. 
However, it is clear from the RGE equations above that large values of $\alpha^{(1)}_2$ lead to a smaller $\beta$ function for $\lambda$ and hence to a slower increase of $\lambda$ at large energies. This behavior is reflected in the plot, with lines of larger $g_1$ including a larger range of $\lambda$ and $\kappa$. Models with 
small $\kappa$ and $\lambda < 1.5$ may remain perturbative consistent up to scales of order of $10^{16}$~GeV.  As shown in sec.~\ref{sec:prec}, this range of $\lambda$'s is also 
preferred for consistency with precision electroweak measurements for $\xi_F = 0$. % For instance, when $g_1$ is 3.5, a combination of $\lambda = 1.4 $ and $\kappa = 0.2 $ would be allowed, which has been guaranteed to give a negative value of the Higgs to $b\bar{b}$ coupling. In order to show this, 

\begin{figure}[H]
    \centering
    \includegraphics[width=16cm]{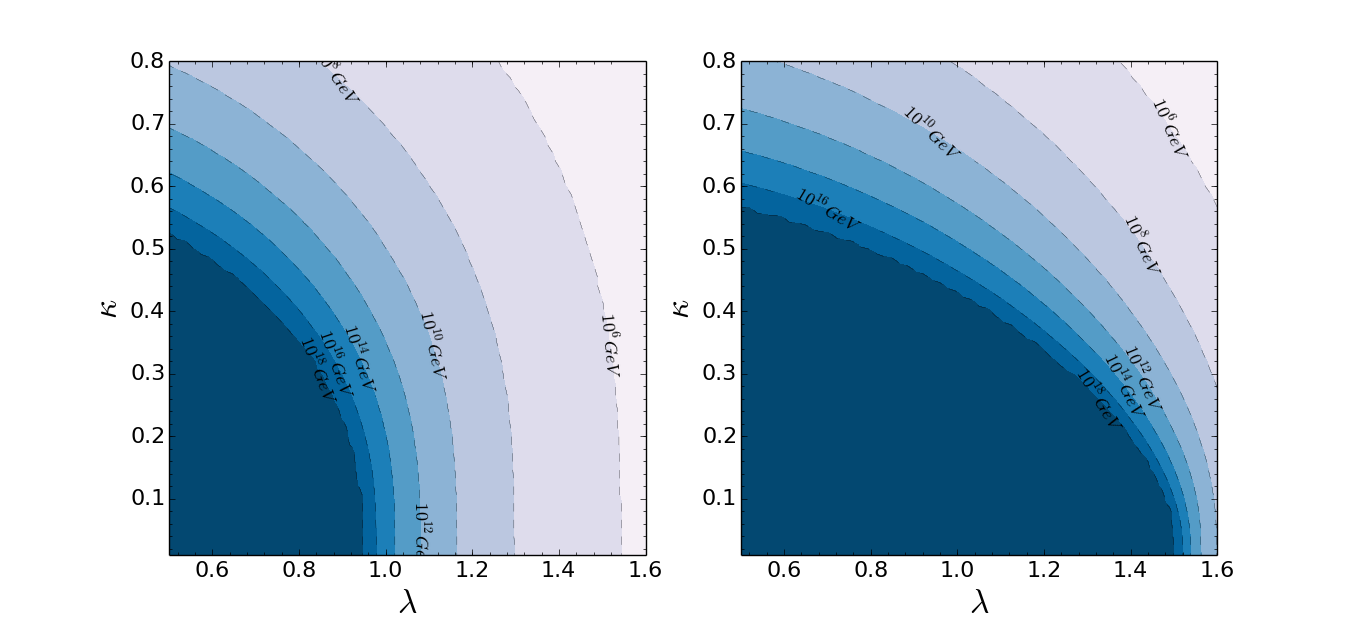}
    \caption{Energy at which some coupling becomes non-perturbative for each $\lambda$ and $\kappa$ combination for a fixed value of $g_1$ at $u = 3$~TeV. The left panel corresponds to $g_1 = 1.5$ while the right panel corresponds to $g_1 = 3.5$. The lines labeled with $10^{16}$ GeV in the two panels are consistent with the two contour lines in Fig.~\ref{new_su2} with the corresponding $g_1$ values.}
    \label{scales}
\end{figure}

In Fig.~\ref{scales} we show the scale at which the coupling $\lambda$ becomes non-perturbative for different weak scale values of $\lambda$ 
and $\kappa$, assuming a symmetry breaking scale $u = 3$~TeV, for two different values of $g_1$ at the scale $u$, namely $g_1 = 1.5$ and $g_1 = 3.5$.  It is
clear that, while for $g_1 = 1.5$ perturbative consistency tends to be lost at scales of the order of $10^6$~GeV, for $g_1 = 3.5$ a wide range of models leading to the inversion of the Higgs coupling to bottom quarks are perturbative consistent up to scales
of the order of $10^{10}$~GeV.

\end{document}